\documentclass[sigconf,authorversion,noacm]{acmart}

\makeatletter
\def\mdseries@tt{m}             
\makeatother                    
\usepackage[draft=true]{minted} 

\usepackage{authblk}
\usepackage{wrapfig}
\usepackage{graphicx}
\usepackage{natbib}
\usepackage{url}
\newcommand\bp{{\bf p}}
\newcommand\bq{{\bf q}}
\newcommand\br{{\bf r}}
\newcommand\bn{{\bf n}}
\newcommand\bI{{\bf I}}
\newcommand\bB{{\bf B}}
\newcommand\bN{{\bf N}}

\newcommand\R{{\mathbb R}}
\newcommand\D{{\mathcal D}}
\newcommand\OO{{\mathcal O}}
\newcommand\mix{\mbox{mix}}
\newcommand\sign{\mbox{sign}}

\newcommand{\secref}[1]{(\S\ref{#1} p.\pageref{#1})}

\newcommand\algif{\mbox{\bf if }}
\newcommand\algthen{\ \mbox{\bf then }}
\newcommand\algelse{\mbox{\bf else }}
\newcommand\algend{\mbox{\bf end }}
\newcommand\algreturn{\mbox{\bf return }}
\newcommand\algand{\ \mbox{\bf and }}
\newcommand\algor{\ \mbox{\bf or }}

\newcommand\algfor{\mbox{\bf for }}
\newcommand\algin{\ \mbox{\bf in }}
\newcommand\algto{\ \mbox{\bf to }}
\newcommand\algwhile{\ \mbox{\bf while }}

\def\new{\textcolor{black}}

\renewcommand\footnotetextcopyrightpermission[1]{} 

\begin{document}
\pagestyle{plain} 

\title{
  Exact predicates, exact constructions and combinatorics \\
  for mesh CSG. \\
  \centerline{\normalsize \rm Bruno Lévy, Inria Saclay, Université Paris Saclay, CNRS, Labo. de Maths. d'Orsay}
  \centerline{\normalsize \tt Bruno.Levy@inria.fr}
  \includegraphics[width=\textwidth]{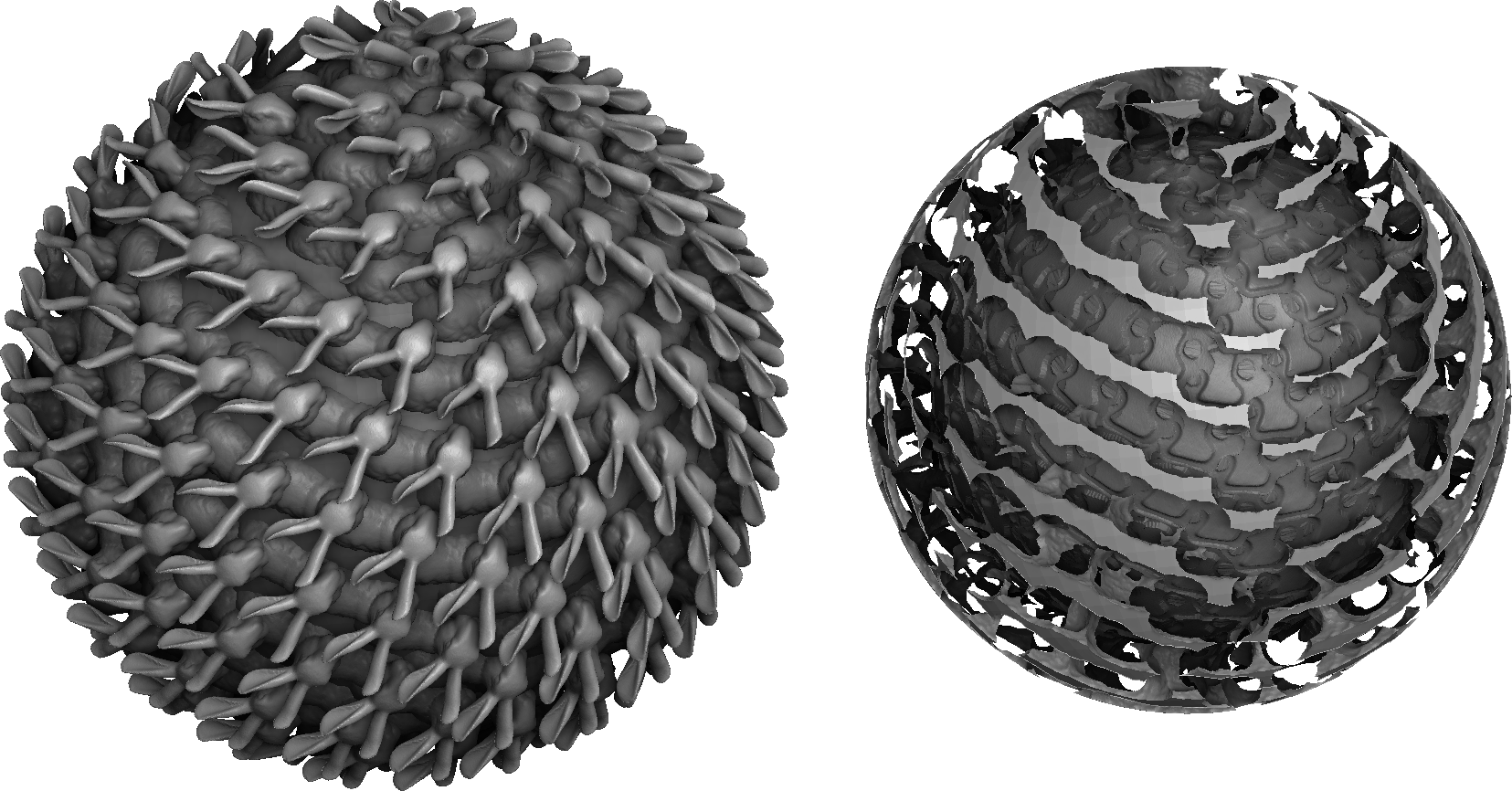}
  {\normalsize \bf  Union and difference between a Fibonacci distribution of
   200 bunnies and a sphere (30 million facets in total).
  }
}

\author[1]{Bruno Lévy}
\affil[1]{Inria Saclay, Université Paris-Saclay, CNRS, Labo. de maths. d'Orsay}



\begin{abstract}
  This article introduces a general mesh intersection algorithm that
  exactly computes the so-called Weiler model \new{(also called a 3D arrangement)}
  and that uses it to implement
  boolean operations with arbitrary multi-operand expressions,
  CSG (constructive solid geometry) and some mesh repair operations.
  From an input polygon soup, the algorithm first
  computes the co-refinement, with an exact representation of the intersection
  points. Then, the decomposition of 3D space into volumetric regions (Weiler model)
  is constructed, by sorting the facets around the non-manifold intersection
  edges (radial sort), using specialized exact predicates.
  Finally, based on the input boolean expression, the triangular facets that belong to
  the boundary of the result are classified.
  \new{The main contribution is a 2D Constrained Delaunay Triangulation with exact coordinates
  that represent the intersections, thanks to two} geometric kernels \new{that} are
  proposed, tested and discussed (arithmetic expansions and multi-precision floating-point).
  As a guiding principle, the combinatorial information shared between each step is
  kept as simple as possible. It is made possible by treating all the particular cases
  in the kernel. In particular, triangles with intersections are remeshed using
  the (uniquely defined) Constrained Delaunay Triangulation, with symbolic perturbations
  to disambiguate configurations with co-cyclic points. It makes it easy to discard
  the duplicated triangles that appear when remeshing overlapping facets.
  The method is tested and compared with previous work, on the existing ``thingi10K''
  dataset (to test co-refinement and mesh repair) and on a new ``thingiCSG''
  dataset made publicly available\footnote{
  as well as the main algorithm and the arithmetic kernel based on expansions,
  see links at the end of the article.} (to test the full CSG pipeline)
  on a variety of interesting examples featuring different types of ``pathologies''.
\end{abstract}

\makeatletter
\g@addto@macro\@maketitle{
  \begin{figure}[H]
  \setlength{\linewidth}{\textwidth}
  \setlength{\hsize}{\textwidth}
  \centering
  \rule{10cm}{5cm} 
  \caption{My first float}
  \end{figure}
}
\makeatother

\maketitle

\section{Introduction and previous work}

\subsection{Why is mesh intersection so hard?}

Mesh intersection is a classical operation in geometry processing. It
is the basic component of higher-level operations, such as
boolean operations, constructing solid geometry, mesh repair, mesh cleaning,
or volumetric modelling operations. However, it is still an important source
of difficulties when implementing geometry processing systems, and it is still an
active area of research and development. At first sight, it may seem rather surprising,
because the problem looks simple:
from a mathematical point of view, what we want to construct is clearly defined.
The input is a set of triangulated surfaces, with possibly intersecting,
or even overlapping/coplanar triangles. The desired output is another set of
triangles, that represent exactly the same surfaces, but that have no intersection.
So why don't we simply ``implement the math''? Why don't we have a standard
implementation that everybody uses? Why is there still active research on this
topic, that is several decades old?
The main difficulty is caused by the representation of coordinates in the
computer. Floating point numbers have a limited precision, which causes two difficulties:

\paragraph*{\bf Geometric predicates: }
  We will need to determine whether a pair of triangles have an intersection, which in turn
  depends on more ``elementary questions'', such as whether a point is above or below a plane
  (in a certain sense). Such
  ``elementary questions'', called \emph{predicates}, are functions that take as an argument a (small)
  number of points (or simple geometric objects) and that returns a set of discrete values.
  For instance, consider four points $\bp_1,\bp_2,\bp_3,\bp_4$. One may want to know the position of $\bp_4$ relative to
  the supporting plane of $\bp_1,\bp_2,\bp_3$, that can be one of {\tt ABOVE}, {\tt BELOW}, {\tt ON\_PLANE}.
  These predicates are the ``nevralgic'' point of mesh intersection methods: if at one moment the
  algorithm ``thinks'' that $\bp_4$ is above the supporting plane of $\bp_1,\bp_2,\bp_3$, it is important that
  at another moment the predicate does not say that $\bp_4$ is below the same plane. How could this happen?
  In fact, these predicates can be expressed as the sign of a polynomial in the coordinates of the points,
  and due to the limited precision of floating point numbers, the output of the predicate can be different
  from the exact mathematical result, especially around zero, and it can depend on the order of the points:
  in floating point arithmetic, imagine you compute $(x_1 + x_2) + x_3$,
  where $x_1 = 1e30$, $x_2 = 1e-6$ and $x_3 = -1e30$ (the result should be $1e-6$). When the
  computer first evaluates $x_1 + x_2$, it gets $1e30$ (because $x_2$ is too small relative to $x_1$), and
  in the end you get $0$. Now if you compute $(x_1 + x_3) + x_2$, you will get a different result ($1e-6$).
  Because of that, you may obtain a different result when you ask for the position of $\bp_4$ relative
  to the supporting plane of $\bp_1, \bp_2, \bp_3$ or when you ask for the position of $\bp_4$ relative
  to $\bp_3, \bp_2, \bp_1$! It can have catastrophic consequences, such as generating an incorrect mesh.

  \begin{figure}
    \centerline{\includegraphics[width=\columnwidth]{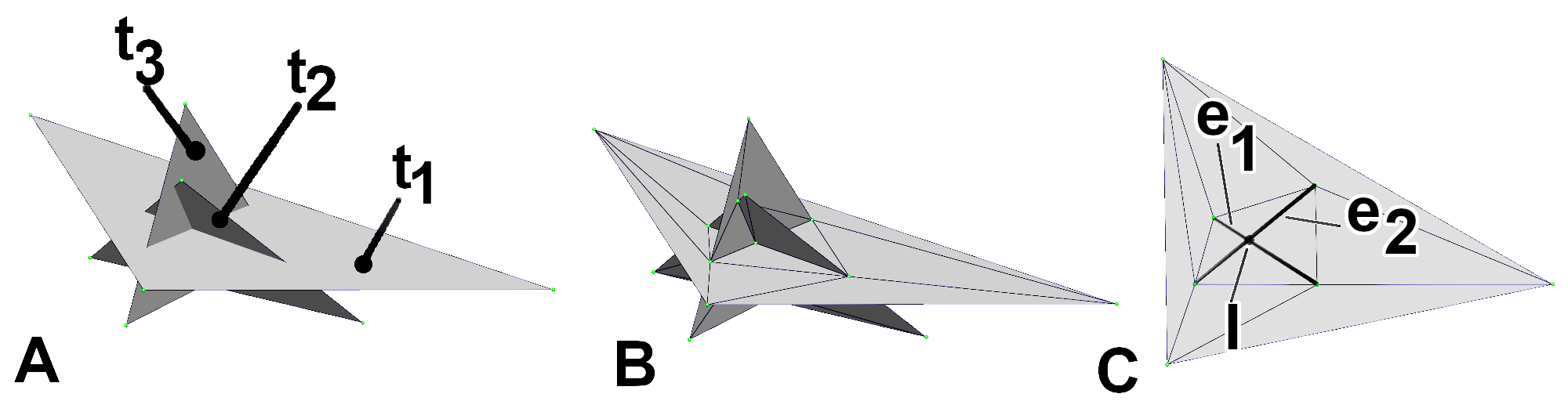}}
    \caption{Intersection between three triangles}
    \label{fig:three_triangles}
  \end{figure}

\paragraph*{\bf Representing intersections: }
  Once the intersecting triangles are determined, one needs to compute the actual intersection. In general,
  the coordinates of the intersection between triangles can be expressed as rational fractions (ratios of
  polynomials in the initial mesh coordinates). Again, in general, they cannot be exactly represented by
  floating-point numbers. It has several consequences: first, imagine you know the exact coordinates of
  all the vertices of your mesh, when you convert them into floating point numbers, they will move a little
  bit. If you do not take care, it may generate new intersections! The problem of constraining the intersection
  points to have floating-point coordinates is referred to as ``snap rounding''.
  It requires special care, that is, one needs to ensure the output with floating-point coordinates does not have
  new intersection and is topologically equivalent (in a certain sense) to the exact result
  \citep{devillers_et_al:LIPIcs:2018:8743}. Second, during the computation, one may need to
  query geometric predicates on the points resulting from intersections, and the answers of these predicates
  need to be coherent with all the rest! For instance, consider a triangle $t_1$ that has some intersections with
  other triangles $t_2, \ldots, t_n$. These intersections create segments $e_1$ and $e_2$ in $t_1$, and one needs to re-triangulate
  $t_1$ in a way that conforms with these segments. In other words, one needs to compute a 2D constrained triangulation
  in $t_1$. This 2D constrained triangulation depends on a set of predicates, and these predicates need to be coherent
  with all the rest. In fact, the situation is even more complicated: as shown in Figure \ref{fig:three_triangles}-A,
  consider that $t_1$ is a triangle that has intersections with two other triangles $t_2$ and $t_3$, mutually intersecting.
  The mesh resulting from the intersection of the three triangles is shown in Figure \ref{fig:three_triangles}-B. When
  remeshing $t_1$, there will be two segments $e_1$ and $e_2$ (highlighted in Figure \ref{fig:three_triangles}-C),
  (one that corresponds to $t_1 \cap t_2$ and the other one to $t_1 \cap t_3$, that create a new intersection $I$).
  Now think of what you have to do to compute this intersection: the extremities of $e_1$ and $e_2$ are intersections between
  input mesh triangles, and we need to compute their intersection. In this specific case, one can compute instead the intersection
  between the supporting planes of the three triangles (that solely depend on the input points), as will be explained later,
  but one needs to keep in mind
  that the intersection $I$ is a constructed points with coordinates
  that are rational fractions, that will be later passed through the geometric predicates
  when computing the constrained triangulation of $t_1$.
  Predicates that depend on constructed points are also used in the subsequent steps of the algorithm,
  such as the ``radial sort'' operation involved in the construction of the Weiler model, that needs new predicates,
  as shall be seen later.

\paragraph*{\bf Software design questions: geometry or combinatorics?}
A mesh intersection / mesh CSG system typically takes the form of a
pipeline composed of several steps (such as \emph{detect candidate triangles intersections,
compute triangle intersections, constrained Delaunay triangulation, merge mesh,
classify intersections, simplify}). There are some impactful decisions to take regarding
the way these steps communicate, in particular, there are two types of information:
\begin{itemize}
\item \emph{geometry}, that is, the exact coordinates of the input points and the constructed intersection points;
\item \emph{combinatorics}, that is, a set of index-based (or pointer-based) data structures that store the connections between
  the triangles (or between some higher-level notions such as charts, shells, regions \ldots).
\end{itemize}

In our context, all point coordinates are exactly represented. As a consequence, the two types of information are redundant:
at any time one could totally reconstruct the combinatorial information from the sole list of triangles and the (exact) coordinates
of their vertices. Hence, the stored combinatorial information either corresponds to the (transient) internal state of one stage
(for instance, a Constrained Delaunay Triangulation), or it is a ``cache'' shared by two stages (or more), ensuring that the
stage(s) downstream do not need to recompute some already known combinatorial information.

Then it would be tempting to always keep all the combinatorial information. However, doing so has the hidden cost of
making the architecture more complicated and more difficult to test. There is a tradeoff to find between a set of independant
and easy-to-test pipeline stages connected by a simple communication protocol, or a fully interconnected set of pipeline stages,
(slightly) more efficient, but (considerably) harder to design, to test and to debug.

\subsection{Summary of this article's contributions}

This article \new{presents}  \new{an algorithm} that computes the so-called
Weiler model \cite{Weiler88} \new{(also called 3D arrangement)} \emph{exactly}, that is, a data structure that
stores the decomposition of 3D space into volumetric regions yielded
by a set of (possibily intersecting) triangulated surfaces.

The algorithm is based on several components. \new{The main new contribution presented here is:}
\begin{itemize}
  \item a new multi-thread friendly constrained Delaunay triangulation algorithm,
  \new{based on revisiting the flip-based algorithm in a way that minimizes
  predicates invocations.}
\end{itemize}

To implement the required exact predicates and exact constructions, two new geometric kernel are described and analyzed:
\begin{itemize}
  \item one is based on arithmetic expansions, like in the approach
    proposed by Shewchuk for predicates \cite{shewchuk96a}, with the difference that arithmetic expansions are also used
    to store constructed points;
  \item the other one is based on multi-precision arithmetics.
\end{itemize}
For both geometric kernels, I explain how to efficiently implement the arithmetic filters,
the predicates, and the symbolic perturbations that ensure the uniqueness of the constrained Delaunay triangulation.
\new{The main benefit of the set of new algorithms introduced here is a significant performance gain (up to 6x as
  compared to \cite{10.1145/3550454.3555460} and \cite{10.1145/2897824.2925901}) in
  degenerate configurations with intersecting co-planar facets (but the integer-based method EMBER \citep{10.1145/3528223.3530181} remains significantly faster). Degenerate co-planar
  configurations are present in some of the models in the Thingi10K database, as shown in the Results section.
  More importantly, such configurations are very often, nearly systematically generated in the CSG trees created py
  practitioners using modeling tools such as the OpenSCAD language, very common in Thingiverse for instance.
}

\new{In the article, to help the practitioner who would want to totally re-implement a similar algorithm,
  I also explain more classical components, also present in the previous work:}
\begin{itemize}
  \item an algorithm to construct the Weiler model;
  \item a (mostly) combinatorial classification algorithm that extracts from the Weiler model the boundary of a region described
    by an arbitrary boolean expression;
  \item an algorithm to simplify the triangulation of co-planar regions. \new{ This algorithm makes use of the
    new constrained Delaunay triangulation, efficiently merging co-planar facets, resulting in simpler triangulations. This
  is especially important when chaining multiple boolean operations present in a deep CSG tree.}
\end{itemize}

The algorithm is tested and compared with previous work on two databases:
\begin{itemize}
  \item The Thingi10K database \cite{Thingi10K}, used to evaluate the co-refinement algorithm;
  \item \new{A small set of scanned meshes and CAD meshes with co-planar facets
    to compare the new method with \cite{10.1145/2897824.2925901} and
    \cite{10.1145/3550454.3555460}};
  \item A new ThingiCSG database with a collection of CSG trees, and the skeleton of an OpenSCAD-compatible
    CSG engine that can be used to test and benchmark future works\footnote{see links at the end of the article.}.
\end{itemize}

This algorithm produces the co-refinement or the result of a boolean operation applied to a set of input meshes, with all
intersection points exactly represented. In the frame of this article, I do not address the (difficult) problem of
converting these exact points into standard floating-point coordinates while preserving some topological
properties (snap rounding). The reader is referred to \cite{devillers_et_al:LIPIcs:2018:8743,valque:hal-02393625} for
an extensive description of snap rounding as well as a possible algorithm.

\subsection{Previous work on mesh intersection}
\label{sec:previous_work}

\paragraph*{\bf Low level (arithmetics)}

At the early times, standard floating point numbers were used, with some carefully tuned threshold and
tolerances to detect corner cases. In the context of tetrahedral meshing, when checking the validity
of a mesh element, it is possible to avoid the arithmetic cancellation problem mentioned in the introduction
by testing all the permutation of the element's vertices. From the early 90's to now, spectacular results were
obtained using the type of strategy mentioned above, and deployed in challenging industrial settings such as highly anisotropic mesh
adaptation for ultrasonic flows \citep{DBLP:journals/impact/GeorgeHS90,DBLP:conf/imr/LoseilleA09}.
Another possibility consists in trying to ``simply implement the math'', in other words, pushing
the difficulties towards the predicates \citep{shewchuk96a,shewchuk97a}, by ensuring that
they exactly follow the definition of the mathematical predicate.  How is it possible with a
computer? Remember, geometric predicates are polynomials in
coordinates of the input mesh's vertices.  The idea in Shewchuk's work
is to use arithmetic expansions, that is, a point coordinate will be
represented by an array of floating point numbers $x_1, x_2, \ldots
x_N$ (instead of a single floating point number). The represented
number corresponds to the sum of all the numbers in the
array. Moreover, these numbers are sorted by decreasing exponents, and
are well separated. That is, the sum $x_2 + x_3 + \ldots x_N$ is
smaller than the floating point value of the least significant bit of
$x_1$. As a consequence, the sign of the represented number is
completely determined by the sign of $x_1$. It is possible to implement
addition, subtraction and multiplication for expansions. If the processor
supports the fused multiply-add instruction {\tt fma} (which is the case of most
modern processors), some noticeable performance gain will be obtained (one of the
basic operations, {\verb|two_product()|} takes 2 instruction with
{\tt fma} versus 13 instructions if {\tt fma} is not available). However, even
with {\tt fma}, operations on expansions are 40 to 100 times slower than with
standard double-precision numbers. For this reason, several strategies were
developed to quickly give the answer in the easy cases. Shewchuk developed
an adaptive precision algorithm, that computes the most significant elements
and refines as needed whenever the sign cannot be determined. However, this
strategy is delicate to
implement\footnote{One may think of an automatic code generator for that.}.

  \begin{figure*}
    \centerline{\includegraphics[width=\textwidth]{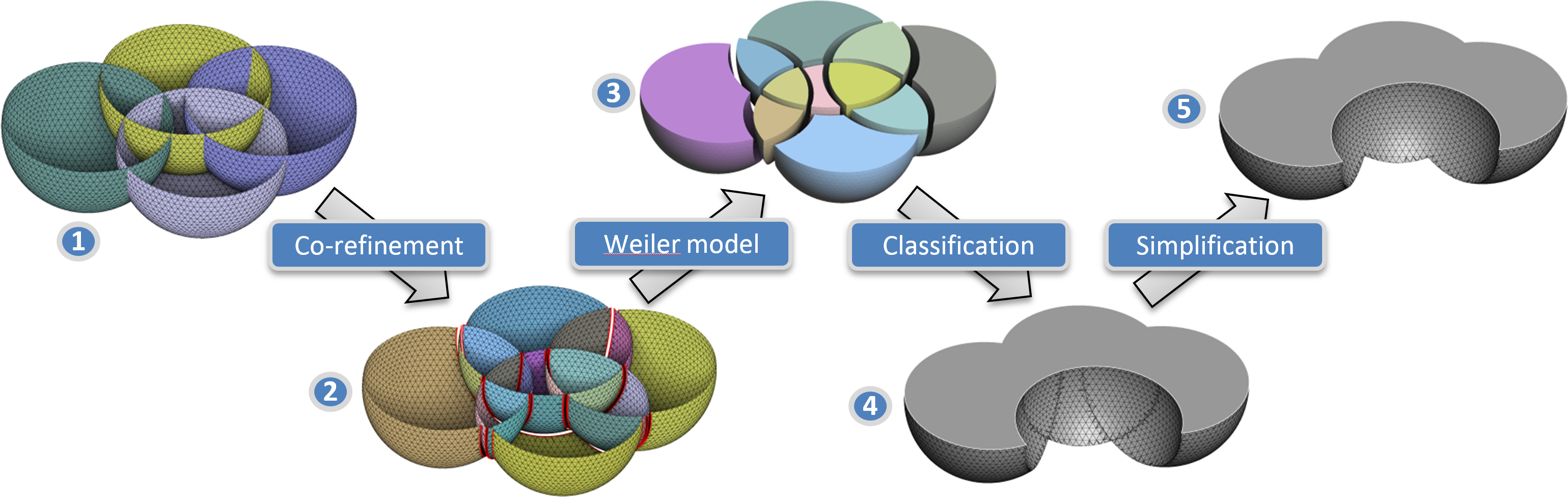}}
    \caption{
      Overview of the algorithmic pipeline for computing boolean operations.
      From a set of intersecting triangulated surfaces (1), we first compute the co-refinement (2), then
      the Weiler model with all volumetric relations (3), from which the result of the boolean operation
      is extracted - here the union of three spheres minus the fourth one (4). Finally, the mesh is
      simplified by merging co-planar triangles (5).}
    \label{fig:CSG_overview}
  \end{figure*}

It is also possible to use interval arithmetics, that is, implementing addition, subtraction and product for
{\tt low},{\tt high} pair of floating point numbers. Each time 0 is not
contained in the [{\tt low},{\tt high}] interval, the sign of the represented
number is known. In all other cases, one needs to relaunch the computation with
expansions. Since it does not happen often, there is a significant performance
gain. Another possibility to quickly determine the sign in the easy configurations
is to use arithmetic filters \citep{meyer:inria-00344297}. The idea shares some similarities with interval
arithmetics, with the difference that it computes with the estimated number
an error bound, using a combination of static information (deduced from the
algebraic expression to be computed) and dynamic information (computed from
the actual values passed to the expression). The \emph{Predicate Construction Kit}
\citep{DBLP:journals/cad/Levy16} takes an algebraic expression of the predicate,
and generates the filter with \emph{FPG} \citep{meyer:inria-00344297}, the
code that computes it with expansions when the filter fails, as well as
symbolic perturbations \citep{DBLP:journals/corr/EdelsbrunnerM94a} for the
degenerate configurations (such as 4 cocyclic points in a Delaunay triangulation). It was used
to generate the non-standard predicates required to compute the intersection between a Voronoi
or a power diagram and a surface or a volume embedded in $n$ dimensions, used in semi-dicrete optimal transport \cite{journals/M2AN/LevyNAL15} and its applications in
fluid simulation \cite{LEVY2022110838} and
cosmology \cite{levy_mnras_2021,vhauss_prl_2022,nikak_prl_2022}.
\new{
  With objectives and ideas similar to the \emph{Predicate Construction Kit}
\citep{DBLP:journals/cad/Levy16} a system for indirect predicates was
introduced \citep{ATTENE2020102856}, with in addition the idea of
storing intermediate construction (that is, new points generated
during the execution of the algorithm)}.
\new{All the techniques mentioned above are based on floating-point numbers. A possible alternative is to use integer numbers only. In \cite{10.1145/3528223.3530181,NehringWirxel2021FastEB}, plane-based representations are used along with homogenous coordinates.
  The impact of this carefully designed integer-based representation is a spectacular performance gain without sacrifying robustness}.

\paragraph*{\bf High Level (intersections and data structures)}
\new{Early works for computing non-exact mesh intersections were proposed
in \cite{douze:hal-01121419}, that proposes a fast algorithm for
mesh intersections used in the context of multi-view acquisition.
The same type of approach was explored in \cite{BARKI20151235}
with special care for robustness, focusing on the involved geometric
predicates}. Using this type of approaches, as well as the techniques for
robust predicates mentioned in the previous paragraph, algorithms and softwares
were developed for robust tetrahedral meshing of triangle soups (that
may have intersections), first in the {\tt TetWild} package
\citep{10.1145/3197517.3201353} that was improved and made more
efficient a while later, based on a more sophisticated algorithm,
dubbed as {\tt FastTetWild}
\citep{10.1145/3386569.3392385}. In the same period of time, the {\tt
  Thingi10K} dataset was published \citep{Thingi10K}. It provides the
research community with a large database of meshes, with many
different cases of degeneracies and inconsistencies. It represents an
excellent stress test for mesh intersection algorithms. \new{For instance,
it was used the same year to test and validate a robust algorithm to
compute the 3D arrangement defined by a set of meshes \cite{10.1145/2897824.2925901}}.

Based on the \emph{indirect predicates} low-level approach mentioned above, a
series of significant advances was published,
first to compute mesh arrangements \citep{10.1145/3414685.3417818} then
boolean operations \citep{10.1145/3550454.3555460}.  Both methods need to compute
constrained triangulations. For that, a highly efficient (linear
complexity) ear cutting algorithm was proposed \citep{9392369}.
The idea of indirect predicates was also used to implement 3D constrained Delaunay triangulation
\citep{10.1145/3618352}, based on an exact representation of the Steiner points.

To help structuring the information through the processing pipeline, several combinatorial data structures
were proposed, such as the Weiler model \citep{DBLP:journals/cga/Weiler85}. This data structure is popular in
computational geosciences, where it is used to represent the volumetric relations between rock layers
and geologic faults \cite{geomodelling,Sword96GM,DBLP:conf/sma/CaumonSM03,PELLERIN201793,DBLP:journals/gandc/LegentilPCFC22}.
\new{There exists exact algorithms for constructing a Weiler model (or 3D mesh arrangement)
  \citep{10.1145/2897824.2925901,10.1145/3414685.3417818,10.1145/3550454.3555460}
and variants that operate on isosurfaces in a volume \citep{10.1145/3528223.3530176}. They
all follow a very similar pipeline:}
they first construct relations between volumes from intersecting surfaces by sorting triangles around non-manifold intersection
edges, then reconstruct the boundaries volumetric zones. To study the combinatorial
aspects of the algorithms, I shall use the notations of combinatorial maps \citep{DBLP:conf/stacs/Lienhardt88}.
Based on an algebraic specification, the notion
of combinatorial maps is well suited to mathematicaly define the operations of a
3D modeler \citep{DBLP:journals/cvgip/BertrandD94}, as well as
multiresolution modeling operators \citep{DBLP:conf/sccg/KraemerCB07} and some operations involved in
hexahedral mesh generation \citep{DBLP:journals/tog/LyonBK16}. In our context, it facilitates describing and implementing
the algorithms in a readable and compact way.

\paragraph*{\bf OpenSCAD and its three geometric kernels} The OpenSCAD software
\citep{WEB:OpenSCAD} is a scriptable CAD package widely used
in the fabrication community. It is based on the CGAL computational
geometry library \citep{WEB:CGAL}, more precisely, it uses the implementation of
Nef complexes \citep{NefComplexes} available in CGAL. To improve computation
time, an alternative algorithm, based on co-refinement, was implemented, also
in CGAL \citep{WEB:CGALcorefinement}. It is significantly faster than the initial
implementation (10x to 50x). All the approaches mentioned above follow the exact geometry
paradigm (exact predicates and exact constructions, except for the final conversion to
floating-point coordinates / snap rounding). There exists a worth mentioning original and highly effective
alternative to this paradigm: instead of doing all the computations in exact mode then converting into
floating-point coordinates while preserving some topological properties, why not doing
all computations in floating point coordinates, while ensuring that each individual computation
preserves some properties? It is the strategy proposed in \cite{ManifoldArticle}, that
builds a network of predicates and operations of increasing dimensions on top of 1D axis-aligned
comparisons (that can be done exactly), while ensuring that the mesh stays manifold at each level.
A fine-tuned parallel implementation was developed \citep{WEB:Manifold} and integrated as an alternative kernel for
OpenSCAD. Since it solely uses single-precision floating point arithmetics, it can also run on the GPU.
The performance gain is spectacular (three orders of magnitude). In this article, I follow the more classical
approach (exact predicates and exact constructions). I implemented an Open-SCAD compatible software to compare
the results with the three geometric kernels mentioned above \secref{sec:thingiCSG}.


\subsection{Overview of this article}

The approach presented in this article shares some similarities with the
works mentioned above, in particular \citep{10.1145/3414685.3417818,10.1145/3550454.3555460,10.1145/2897824.2925901},
with the following differences:
\begin{enumerate}
\item \new{unlike in \cite{10.1145/3414685.3417818,10.1145/3550454.3555460}
  exact constructions are used instead of Predicate Construction Kit or indirect predicates.
  Note that exact constructions (from CGAL) are also used in \cite{10.1145/2897824.2925901}. We
    further investigate the geometric kernel with two new numbers representations and the associated
    predicates, for significant gains (up to 6x) in degenerate configurations, as compared to the
    previous works mentioned above};
  \item the algorithm computes a constrained \emph{Delaunay}
    triangulation (CDT), which ensures the uniquenes of the obtained triangulation. \new{This comes
      as a higher computational cost, related to the} \verb|in_circle()| \new{ predicate, as compared to
      \cite{10.1145/3414685.3417818} that computes a (non-Delaunay) constrained triangulation.}
    The uniqueness of the triangulation is useful to make
    sure that the same triangles will be generated in configurations with co-planar overlapping facets.
    Then duplicated triangles can be easily eliminated, whereas previous works
     need to use an auxilliary data structure to identify overlapping zones.
     Note that the \verb|in_circle| predicates depends on 4 points, that can have different
     expressions \new{(they can be input data points, or intersections of different types)}. Using
    Predicate Construction Kit or indirect predicates would generate a large number of
    instances of the \verb|in_circle| predicates (depending on whether the points are
    initial vertices or intersections). Here this ``combinatorial explosion'' is avoided thanks
    to the exact constructions. In a certain sense, the complexity is pushed towards the kernel.
    \new{Another benefit of this efficient CDT algorithm is the ability of efficiently re-triangulating
      zones made of co-planar triangles. This is expecially important when chaining boolean operations, as
      in the evaluation of CSG trees, to avoid generating a huge number of triangles};
  \item the algorithm computes a 3D partition of space into volumes, \new{that is, a 3D mesh arrangement},
    represented in a data structure called the
 Weiler model \citep{DBLP:journals/cga/Weiler85}.
 I shall explain later how this data structure can be efficiently constructed, and how most of the calls to the
 predicates can be avoided by exploiting the combinatorics. Then I shall use this data structure
 to implement boolean operations and CSG primitives, still exploiting the combinatorics as much as possible.
 \new{This is very similar to the 3D mesh arrangement construction in \cite{10.1145/2897824.2925901}. The difference
   is that I describe it using the notion of combinatorial map \cite{DBLP:journals/ijcga/Lienhardt94}, that
   yields a compact form of the combinatorial operations such as the classification.}
\end{enumerate}

This article introduces an algorithmic pipeline depicted in Figure \ref{fig:CSG_overview}.
Given an input polygon soup, the algorithm computes the co-re\-fi\-ne\-ment
\secref{sec:corefinement}, based on an exact constrained Delaunay triangulation
constructed in each facet that has intersections \secref{sec:CDT2d}.
Then, the algorithm creates the Weiler model \secref{sec:Weiler},
that represents all the radial relations around the
non-manifold intersection edges as well as the volumetric relations
between the regions delimited by the surfaces.  Finally, from the
input boolean expression, a classification algorithm selects the
triangular facets that belong to the boundary of the result \secref{sec:classify}.
In addition, an optional step merges and re-triangulates co-planar
facets \secref{sec:simplify}.
To exactly represent the intersection points, I implemented
and tested two different arithmetic kernels \secref{sec:arithmetics}.
One is based on arithmetic expansions \secref{sec:expansions},
and the other one on multi-precision floating point arithmetics
\secref{sec:mpfloat}. Both are pre-filtered using interval arithmetics.  The
uniqueness of the Delaunay triangulation is ensured, even in
configurations with cocyclic points, by using symbolic
perturbations. This ensures that regions overlapped by multiple
coplanar input facets are coherently and uniquely triangulated.  Finally,
I tested the method and compared with previous work \secref{sec:tests},
on the existing ``thingi10K'' dataset for the co-refinement algorithm,
\secref{sec:thingi10K} and on a new
``thingiCSG'' dataset together with an OpenSCAD compatible CSG engine, both
made publicly available\footnote{both publicly available,
as well as the main algorithm and the
arithmetic kernel based on expansions, see links at the end of the article.},
to test the full CSG pipeline
on a variety of interesting examples featuring different types of
``pathologies'' \secref{sec:thingiCSG}.

  \begin{figure*}
    \centerline{\includegraphics[width=\textwidth]{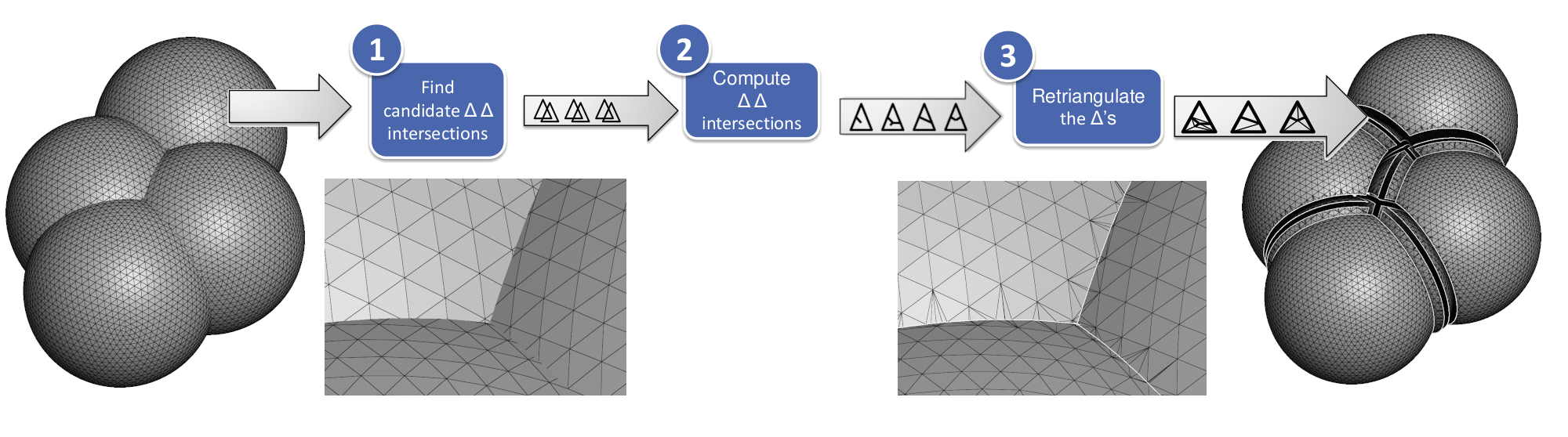}}
    \caption{Surface mesh co-refinement takes as input a set of triangulated surface (left). Then, the
      candidate pairs of intersecting triangles are determined (1), which generates a stream of pairs of
      triangles. Triangles intersections are computed (2). The result is a stream of triangles with the set
      of segments that would be inserted into each of them. Finally, a constrained Delaunay triangulation
      is computed in each triangle (3) and the resulting triangles are merged to create the co-refinement mesh
      (right, displayed in ``exploded view'').}
    \label{fig:corefinement}
  \end{figure*}

\section{The algorithm}

\subsection{Mesh co-refinement}
\label{sec:corefinement}

Mesh co-refinement takes as input a set of triangulated surfaces. No assumption is made regarding the structure of this input.
It can be a triangle soup made of disconnected triangles. It can have co-planar facets. Duplicated vertices are pre-detected
and merged using \new{lexicographic} sorting. Duplicated facets are then pre-detected and discarded, using lexicographic sort. Then, the
different substeps of the algorithm, outlined in Figure \ref{fig:corefinement}, are as follows:

\begin{itemize}
  \item Detect candidate triangle intersection pairs \secref{sec:candidates};
  \item Compute triangle intersections \secref{sec:tritri};
  \item Re-triangulate the intersected triangles, using a constrained Delaunay triangulation \secref{sec:CDT2d}
\end{itemize}

The output is a valid mesh that exactly represents the same geometry as the input mesh. It is valid in the sense it has
no intersection. Output vertices stemming from triangle intersections are exactly represented. We shall see two different
ways of doing that, using either arithmetic expansions or multi-precision floating-point numbers  \secref{sec:arithmetics}.

\subsubsection{Detecting candidate intersecting facet pairs}
\label{sec:candidates}

The first step of the algorithm determines all pair of potentially intersecting facets. To avoid having to test
the $N_F^2/2$ pairs of triangles, we use the classical AABB data structure (axis-aligned bounding box tree).
Readers already familiar with AABBs may skip this subsection. However, I found it useful to give here some details
and references, in particular about the idea that an AABB can be stored in compact form, mostly by re-ordering the mesh
elements, as done in the OpCode library \cite{ZBAABB1} (and in more recent ones, such as PhysX).
An AABB is a set of nested boxes, organized in a binary tree. Each internal node contains two children
(each of them having its own bounding box),
and in my implementation, each leaf contains a single triangle.
Each internal node $n$, encoded as an integer, knows its two children and
its bounding box. The two children are implicitly encoded, by
$\mbox{left\_child}(n) = 2n$ and $\mbox{right\_child}(n) = 2n+1$ (note that the root node needs to be $1$ rather than
$0$, else the root would be its own left child!). The bounding boxes are simply stored in a contiguous array
indexed by $n$. Now we need to know which triangles correspond to a given node. Again, we store as few information
as possible. Instead of storing a set of triangle index in each node, we permute the mesh triangles in such a way
that all the triangles corresponding to a given node are contiguous in memory. Hence, the entire mesh, or the
sequence of triangle indices $[0, \ldots N_F)$ corresponds to the root node $n=1$. Then the sequence
$[0, \ldots N_F/2)$ corresponds to its left child $n=2$, and the sequence $[N_F/2, \ldots N_F)$ to its right
child, and so on and so forth. Hence node $n$ contains the sequence of triangles $[b,e)$,
$\mbox{left\_child}(n)$ (with index $2 \times n$) contains the sequence $[b,m)$ and
$\mbox{right\_child}(n)$ (with index $2 \times n+1$) contains the sequence $[m,e)$, with $m = b+(e-b)/2$. With this
encoding, instead of being stored, triangle indices are implicitly determined, and propagated through
the recursive function calls that traverse the tree. \\

Such a AABB tree can be easily constructed by first re-ordering the mesh facets, then computing the bounding boxes
of each facet, and then recursively create the bounding boxes of higher level nodes. To re-order the mesh facets,
we use balanced Morton ordering (also called balanced Z-curve ordering). It can be easily implemented using the \verb|std::nth_element|
function of the Standard Template Library, as done in the spatial sorting package of the CGAL library \citep{WEB:SpatialSorting}.
The re-ordering of the facets is computed in-place in the mesh data structure. Since we used a balanced AABB-tree, the links
in the tree are completely implict. The only thing that needs to be stored is the array of bounding boxes. Depending on the
importance of performance w.r.t. storage requirement, other choices are possible, such as using unbalanced trees constructed
with the classical Surface Area heuristic (SAH) \cite{10.1007/BF01911006}. Unbalanced trees require storing for each node the
number of triangles in one of the subtrees. On the other end of the spectrum, zero-byte AABBs \cite{ZBAABB1,ZBAABB2}, do not
need any additional data structure. They are based on the observation that the bounds of bounding box are coordinates of
some vertices in the mesh. At the expense of a small number of additional tests, this makes it possible to encode the
entire acceleration structure within the ordering of the triangles and their vertices. In our context, balanced AABBs realize
a good compromise between speed and storage.

Now, given a AABB tree that contains all the facets to be intersected, we need to write a function that will determine a superset
of all pairs of facets that have an intersection. This superset will correspond to the pairs of facets which bounding boxes have
an intersection. It is based on a recursive AABB tree traversal, that determines all the intersections between all bounding boxes
in two subtrees planted at nodes $n_1$ and $n_2$ respectively. Node $n_1$ corresponds to the facets sequence $[b_1,e_1)$ and node
$n_2$ to facets sequence $[b_2,e_2)$. The function also takes as an argument a function $\mbox{DO\_IT}$, that will be called
for each candidate facet pair:

$$
\begin{array}{ll}
  \mbox{\bf input:} \\
  \quad n_1, b_1, e_1: & \mbox{first node } (n_1) \mbox{ and associated facet sequence } [b_1,e_1) \\
  \quad n_2, b_2, e_2: & \mbox{second node } (n_2) \mbox{ and facet sequence } [b_2,e_2) \\
    \quad \mbox{DO\_IT}: & \mbox{ function to be called for candidate intersecting facets } \\
    \hline \\
  (1)  & \mbox{intersect}(n_1, b_1, e_1, n_2, b_2, e_2, \mbox{DO\_IT}) \\
  (2)  & \quad \algif e_2 \le b_1 \algthen \algreturn \\
  (3)  & \quad \algif \mbox{bbox}[n_1] \cap \mbox{bbox}[n_2] = \emptyset \algthen \algreturn \\
  (4)  & \quad N_1 \leftarrow e_1 - b_1 \quad ; \quad N_2 \leftarrow e_2 - b_2 \\
  (5)  & \quad \algif N_1 = 1 \algand N_2 = 1 \algthen \mbox{DO\_IT}(b_1, b_2) \\
  (6)  & \quad \algif N_2 > N_1 \algthen \\
  (7)  & \quad \quad m_2 \leftarrow b_2 + N_2/2 \\
  (8)  & \quad \quad \mbox{intersect}(n_1, b_1, e_1, \mbox{left\_child}(n_2), b_2, m_2) \\
  (9)  & \quad \quad \mbox{intersect}(n_1, b_1, e_1, \mbox{right\_child}(n_2), m_2, e_2) \\
  (10) & \quad \algelse \\
  (11) & \quad \quad m_1 \leftarrow b_1 + N_1/2 \\
  (12) & \quad \quad \mbox{intersect}(\mbox{left\_child}(n_1), b_1, m_1, n_2, b_2, e_2) \\
  (13) & \quad \quad \mbox{intersect}(\mbox{right\_child}(n_1), m_1, e_1, n_2, b_2, e_2) \\
  (14) & \quad \algend \\
  (15) & \algend
\end{array}
$$

Algorithmic details:
\begin{itemize}
\item Line $(2)$ exits the function if the facet sequence corresponding to $n_2$ is ``to the left''
  of the one corresponding to $n_1$. This avoids doing the same traversals twice (once
  with $n_1,n_2$ and once with $n_2,n_1$);
\item Line $(3)$ early exits the function if the two bounding boxes of $n_1$ and $n_2$ are non-overlapping.
  It is where the acceleration occurs;
\item Line $(4)$ computes the number of facets $N_1$ in $n_1$ and $N_2$ in $n_2$;
\item Line $(5)$ handles leaf-leaf intersections, by calling $\mbox{DO\_IT}$;
\item Lines $(7)-(14)$ recursively compute intersections by traversing the children
       of the node that has the largest number of triangles.
\end{itemize}

At the top level, recursion is launched by calling
intersect$(1,0,N_F,$ $1,0,N_F)$ where $1$ corresponds to the root
node of the AABB, and where $N_F$ denotes the number of facets in the
mesh.

\subsubsection{Computing the intersection between two triangles}
\label{sec:tritri}

  \begin{figure*}
    \centerline{\includegraphics[width=\textwidth]{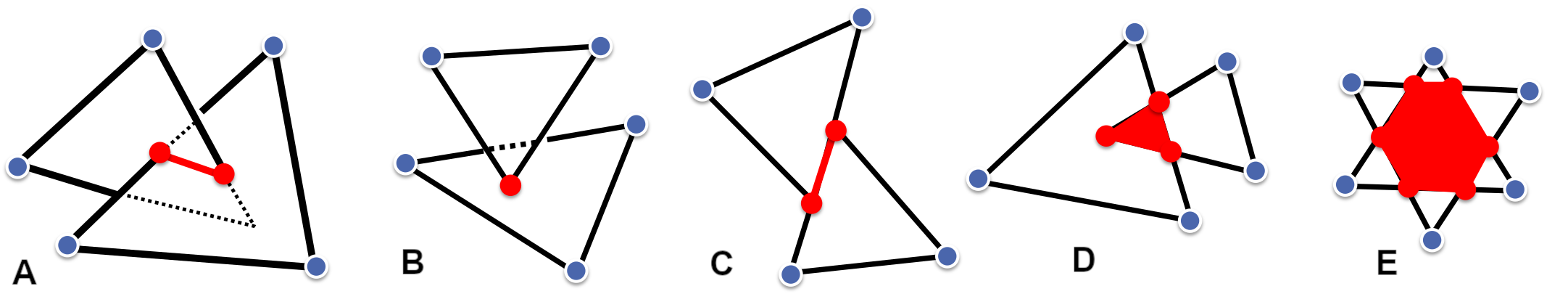}}
    \caption{Intersection between two triangles, a surprisingly delicate problem.
      Some of the configurations one may encounter. There are many other ones!}
    \label{fig:tritri}
  \end{figure*}

The output of the previous step of the algorithm is a stream of
potentially intersecting triangle pairs $(t,t^\prime)$, generated by
the calls to the $\mbox{DO\_IT}(t,t^\prime)$ function. Now we need to determine within these
candidates the ones that correspond to actual intersections. In
addition, we need also to determine the coordinates of the
intersection points. At first sight, computing the intersection
between two triangles is a rather simple task.  Whether two triangles
overlap can be \emph{exactly} determined using the \verb|orient_3d()| predicate
\citep{devillers:inria-00072100}.  But in our case, we need more
information. We need not only the coordinates of the intersection, but
also the associated combinatorial information. Moreover, there is a
(surprisingly) large number of possible configurations for the
intersection of two triangles. Figure \ref{fig:tritri}-A shows the
generic case, where the intersection between the two triangles is a
line segment. Each extremity of this line segment corresponds to the
intersection between one of the triangle's edges with the supporting
plane of the other triangle. However, there are many possible degeneracies,
such as a point of one triangle that falls exactly on the other one (B),
or two edges that partially overlap (C). If the triangles are coplanar,
the intersection can even be a polygon (D) with up to 6 vertices (E)! \\

\begin{figure}[b]
  \centerline{
    \includegraphics[width=0.75\columnwidth]{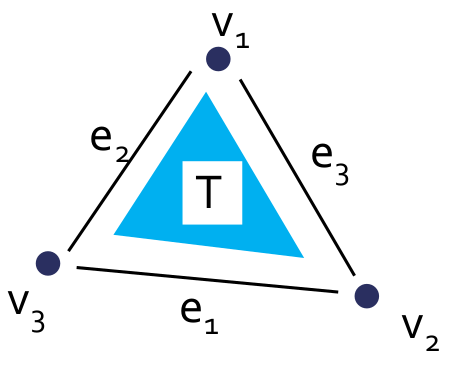}
  }
  \vspace{10mm}
  \centerline{
    {\Huge
      $\Sigma_t = \{ v_1, v_2, v_3, e_1, e_2, e_3, T\}$
    }
  }
  \vspace{5mm}
  \caption{
    A triangle $t$ seen as a simplicial set $\Sigma_t$, decomposed into 7 open regions:
    its three vertices $v_1$, $v_2$, $v_3$, the three edges $e_1$, $e_2$, $e_3$ (minus their extremities)
    and the ``meat'' of the triangle $T$ (minus the border). This (trivial)
    way of considering a triangle makes it easy to design a correct algorithm that determines the combinatorial
    intersection between two triangles that works for all possible degenerate cases (some of
    them depicted in Fig. \ref{fig:tritri} above).
  }
  \label{fig:tri_simplices}
\end{figure}

To tackle this problem, we will write a function that computes a \emph{combinatorial}
representation of the intersection. Each triangle $t$ (resp. $t^\prime$) can be
seen as a simplicial set $\Sigma_t$ with 7 simplices: each triangle has three vertices $V_1, V_2, V_3$,
three edges $E_1 = (P_2, P_3)$, $E_2 = (P_3, P_1)$, $E_3 = (P_1, P_2)$, and the whole triangle
$T = (P_1, P_2, P_3)$.

We consider that each simplex is embedded as an open set. In other words,
each edge is embedded as a segment minus the extremities, and $T$ is embedded
as the triangle minus its border, hence each point of the triangle is contained by exactly one simplex.
For each intersection point (in red in Figure \ref{fig:tritri}) we will output the unique pair of
simplices $\sigma, \sigma^\prime$ whose embeddings contain the point, where $\sigma \in \Sigma_t = \{ P_1, P_2, P_3, E_1, E_2, E_3, T \}$
and $\sigma^\prime \in \Sigma_{t^\prime} = \{ P_1^\prime, P_2^\prime, P_3^\prime, E_1^\prime, E_2^\prime, E_3^\prime, T^\prime \}$.
In practice, in an implementation, the simplices can be encoded as integers, or enums.
From this description, one can imagine the following (generic, naive) triangle-triangle intersection algorithm, that tests
all the $7 \times 7$ possible intersections between the simplices of $t$ and $t^\prime$:

$$
\begin{array}{ll}
  &\mbox{triangle\_triangle\_naive()} \\
  &\mbox{\bf input:}
   \quad \mbox{two triangles } t \mbox{ and } t^\prime \\
  &\mbox{\bf output:}
   \quad \mbox{a list } I \mbox{ of simplices pairs } (\sigma,\sigma^\prime) \\
   & \quad \quad \mbox{ that define the intersection points} \\
  \hline
  \\
  (1) & \algfor \sigma \algin \Sigma_t \\
  (2) & \quad \algfor \sigma^\prime \algin \Sigma_{t^\prime} \\
  (3) & \quad \quad \algif \sigma \cap \sigma^\prime \mbox{ is a point } \algthen \\
  (4) & \quad \quad \quad I \leftarrow I \cup \{ (\sigma, \sigma^\prime) \} \\
  (5) & \quad \quad \algend \\
  (6) & \quad \algend \\
  (7) & \algend \\
  (8) & \algreturn I
\end{array}
$$

It is possible to be much smarter than what is done in the algorithm above. For instance, the triangle-triangle intersection test
in \cite{devillers:inria-00072100} determines whether two triangles have an intersection, and minimizes the number of
\verb|orient3d| invocations. In our case, we cannot use it directly, because our situation is slightly more complicated:
we also need to compute the combinatorial representation of the intersection. But we can avoid some unnecessary tests as follows:

$$
\begin{array}{ll}
  &\mbox{triangle\_triangle()} \\
  &\mbox{\bf input:}
   \quad \mbox{two triangles } t \mbox{ and } t^\prime \\
  &\mbox{\bf output:}
   \quad \mbox{a list } I \mbox{ of simplices pairs } (\sigma,\sigma^\prime) \\
   & \quad \quad \mbox{ that define the intersection points} \\
  \hline
  \\
  (1) & \algif P_1, P_2, P_3 \mbox{ are strictly on the same side } \\
      & \quad \quad \mbox{of the support plane of } (P^\prime_1, P^\prime_2, P^\prime_3) \algthen \algreturn \emptyset\\
  (2) & \algfor E \algin \{ E_1, E_2, E_3 \} \\
  (3) & \quad I \leftarrow I \cup \mbox{edge\_triangle}(E,T^\prime) \\
  (4) & \algend \\
  (5) & \algfor E^\prime \algin \{ E^\prime_1, E^\prime_2, E^\prime_3 \} \\
  (6) & \quad I \leftarrow I \cup \mbox{edge\_triangle}(E^\prime,T) \\
  (7) & \algend \\
  (8) & \algreturn I
\end{array}
$$

The idea is to have an early-exit test (line 1), then test the three edge of each triangle against the other one. The
underlying edge-triangle intersection algorithm works as follows:

$$
\begin{array}{ll}
  &\mbox{edge\_triangle()} \\
  &\mbox{\bf input:}
  \quad \mbox{an edge } E \mbox{ and a triangle } T \\
  & \quad \mbox{\new{(both considered as closed sets, with extremities and edges)}} \\
  &\mbox{\bf output:}
   \quad \mbox{a list } I \mbox{ of simplices pairs } (\sigma,\sigma^\prime) \\
   & \quad \quad \mbox{ that define the intersection points} \\
  \hline
  \\
  (1) & (P_1, P_2, P_3) \leftarrow \mbox{vertices}(T) \\
  (2) & (Q_1, Q_2) \leftarrow \mbox{vertices}(E) \\
  (3) & s_1 \leftarrow \mbox{orient3d}(P_1,P_2,P_3,Q_1) \\
  (4) & s_2 \leftarrow \mbox{orient3d}(P_1,P_2,P_3,Q_2) \\
  (5) & \algif s_1 \times s_2 > 0 \algthen \algreturn \emptyset \\
  (6) & \algif s_1 = 0 \algand s_2 = 0 \algthen \algreturn \mbox{edge\_triangle\_2D}(E,T) \\
  (7) & o_1 \leftarrow \mbox{orient3d}(Q_1,Q_2,P_1,P_2) \\
  (8) & o_2 \leftarrow \mbox{orient3d}(Q_1,Q_2,P_2,P_3) \\
  (9) & o_3 \leftarrow \mbox{orient3d}(Q_1,Q_2,P_3,P_1) \\
  (10) & \algif o_1 \times o_2 < 0 \algor o_2 \times o_3 < 0 \algor o_3 \times o_1 < 0 \algthen \algreturn \emptyset \\
  (11) & R_1 \leftarrow \mbox{region}(T,o_1,o_2,o_3) \\
  (12) & R_2 \leftarrow \mbox{region}(E,s_1,s_2) \\
  (13) & \algreturn (R_2,R_1)
\end{array}
$$

Algorithmic details:
\begin{itemize}
  \item Line $(5)$: if both extremities of $E$ are on the same side of $T$ then there is no intersection;
  \item Line $(6)$: if both extremities of $E$ are on the supporting plane of $T$ then we are in 2D and it
     is a special case, using a different codepath (more on this later);
  \item Line $(10)$: if two of $o_1$, $o_2$, $o_3$ have opposite sign, then the intersection between $E$ and
    the supporting plane of $T$ is outside $T$;
  \item Line $(11)$: if exactly one of $o_1$, $o_2$, $o_3$ is 0, then the intersection is on an edge of $T$. If
    two of them are 0, then the intersection is on a vertex of $T$. The function $\mbox{region}$ returns this
    edge or this vertex or $T$ depending on which of $o_1$, $o_2$, $o_3$ is zero;
  \item Line $(12)$: same thing with the edge $E$: if one of $s_1$, $s_2$ is zero, then the intersection is on
    a vertex of $E$;
\end{itemize}

The function $\mbox{edge\_triangle\_2D}()$ will not be detailed explicitly here\footnote{but is available in the companion source-code, see links at the end of the article.},
but we give an idea of how it works. It first finds
a direction of projection, then determines whether, in 2D, $E$'s extremities are in $T$, then computes the
intersection between $E$ and the three edges of the triangle in 2D. For edges that are co-linear (like in
Figure \ref{fig:tritri}-C), there is a $\mbox{edge\_triangle\_1D}()$ function that determines the intersection
by comparing intervals. \\

The number of invocations to the \verb|orient_3d| predicate remains higher than the optimal, because our algorithm naively computes
intersection between sub-simplices without taking into account more global information. To avoid doing the same computations several
times, I use a ``predicate cache'' \new{(a similar technique is used in \cite{10.1145/3550454.3555460})}. A predicate cache is a table that maps predicate's input to the resulting sign. The key used to index the table is the list of the sorted indices of the input points. When querying the predicate, if present in the table,
  the stored sign is flipped depending on the parity
of the actual order of the four arguments (\verb|orient_3d| is a determinant, hence permutting its columns changes the sign depending
on the parity of the permutation). \\

The output of the algorithm is a list $I$ of couples $(\sigma,
\sigma^\prime)$ that correspond to all intersection points (red dots
in Figure \ref{fig:tritri}). Each simplex $\sigma$, $\sigma^\prime$
can be encoded as an integer in $[0,\ldots 6]$. Note that since we
divided the initial problem (triangle-triangle intersection) into
simpler independent problems (edge-triangle, edge-edge intersections),
one may obtain the same intersection several times. Duplicated
intersections can be eliminated by sorting the list of couples
$(\sigma, \sigma^\prime)$ (using \verb|std::sort()| with the
lexicographic order for instance) and eliminating the duplicates
(using \verb|std::uniq()|).

The method we have described so far is able to handle triangle-triangle
intersection when the result is a segment (Figure \ref{fig:tritri}-A and C)
or a point (Figure \ref{fig:tritri}-B). When the intersection is a polygon
(Figure \ref{fig:tritri}-D and E), the method outputs all the vertices of
the polygon, but they come in an arbitrary order, so we need to order them
along the boundary of the intersection polygon. In fact, what we need is
finding all the edges of the intersection polygon (because they will be
positioned as Delaunay constraints, more on this in the next subsection).
The idea is to test all possible edges of the intersection polygon, and keep
only the ones that are on the boundary of the intersection polygon (red edges
in Figure \ref{fig:tritri} D and E). In other words, given two intersection points,
defined by the couples of simplices $(\sigma_1,\sigma^\prime_1)$ and $(\sigma_2,\sigma^\prime_2)$,
how can we determine if the so-defined edge is on the border of the intersection?
One can observe that an edge of the intersection is always a subset of an edge of one of
the triangles, hence we just need to test if $\sigma_1$ and $\sigma_2$ are on the same edge of $t$
or $\sigma^\prime_1$ and $\sigma^\prime_2$ are on the same edge of
$t^\prime$. This test is very simple, two simplices $\sigma_1$ and $\sigma_2$ are on the same edge if:
\begin{itemize}
 \item $\sigma_1 = \sigma_2 = e$ where $e$ is an edge, or
 \item $\sigma_1$ is an edge and $\sigma_2$ is a vertex of $\sigma_1$, or
 \item $\sigma_2$ is an edge and $\sigma_1$ is a vertex of $\sigma_2$
\end{itemize}

To summarize, once we have determined all the intersection points,
there are three cases:
\begin{enumerate}
  \item{\bf there is a single point:} the intersection is degenerate and is a point;
  \item{\bf there are two points:} the intersection is the segment that connects both points;
  \item{\bf there are more than two points:} the intersection is a polygon. One obtains its edges by testing all possible
    couples of points (maximum 15 couples to test). It is a trivial combinatorial test.
\end{enumerate}

At the end of this step, what we obtain is for each triangle $t$, a list of segments generated from an intersection
between $t$ and other triangles. For each segment extremity, we know the couple of simplices $\sigma,\sigma^\prime$
that generated the intersection. The next step of the algorithm is to remesh each triangle in such a way that
all segments are explicitly represented in the resulting mesh.

\subsubsection{Constrained Delaunay Triangulation}
\label{sec:CDT2d}

We need to insert a list of points and segments in each triangle, hence we need to compute a large number of
\emph{constrained} triangulations. These triangulations are computed in 2D. It is possible to use 2D
coordinates in the supporting plane of each triangle, however, with the exact number representation that we use, it has
the non-negligible cost of nesting additional dot products in the expressions. As often done in other works,
I chose instead to peek the two coordinates of the triangle's normal that have the smallest absolute values.
As noted in \cite{10.1145/3414685.3417818}, this requires special care:
comparing the magnitudes of the components of the normal vector needs to be done in exact arithmetics, else one may project
along axes that create degenerate configurations. It can happen for instance with a very skinny triangle with a normal
close to $[1,1,1]$. Due to floating point rounding error, one may pick a projection axis onto which the triangle degenerates
into a segment. Invoking exact arithmetics for finding the dominant axis of a normal vector is not pedantic as one may
think: I encountered this problem with mesh \verb|#356074| from \cite{Thingi10K}\footnote{I find it worth it to confess
that I learned the lesson the hard way, required some debugging, I should have believed them \citep{10.1145/3414685.3417818}
right from the beginning!}.

In addition to the triangulation being \emph{constrained}, it is interesting to require it to be a constrained \emph{Delaunay}
triangulation for two reasons:

\begin{itemize}
\item With Delaunay, the quality of the mesh is ``not too bad'', because it maximizes the smallest angle. I say here ``not too
  bad'' because in general, even with Delaunay, intersection meshes contain small angles and cannot be directly used in numerical
  simulation without some re-meshing / post-processing. However, it is always good to have a starting point that does not have
  too many triangles with very small angles;
\item the Delaunay triangulation is unique, which is an interesting property when the intersecting meshes have overlapping coplanar
  facets. This property ensures that the same zone will be meshed with the same triangles (one only needs to filter-out the duplicated
  triangles). There exists other method to solve the problem with non-unique triangulations, based on a cavity-remeshing operator
  with linear complexity \citep{9392369}, but this require maintaining a list of
  polygons together with the triangulations, and identifying the duplicated polygons, using a more complicated data structure.
  In other words, this means pushing the difficulties into the combinatorial data structure. I
  prefer to keep them in the predicates, because predicates are concentrated in a small portion of the code, easier to maintain
  and to debug.
\end{itemize}

In our ``wish list'' for the constrained Delaunay triangulation code, we need the following two properties:

\begin{itemize}
\item \textbf{genericity:} since the extremities of the constrained edges are intersection points, their coordinates
  are not represented as standard floating-point, and the algorithm needs to be adapted to these ``exotic'' points;
\item \textbf{efficiency:} the new code will be deployed in an industrial context, and needs to have performances that
  are on par with the standards. In particular, in principle, re-meshing all pairs of intersected triangles can be performed in
  parallel, so the constrained Delaunay triangulation code needs to be multithread-friendly (no dynamic allocations, as few
  locks as possible).
\end{itemize}

There are several implementations of a constrained Delaunay
triangulation available, such as Shewchuk's \emph{Triangle}
\citep{shewchuk96b} and CGAL \citep{WEB:CGAL}. I chose not to
use them for several reasons. First, \emph{Triangle} has hardwired predicates (whereas we need to plug special ones,
adapted to points that come from intersections), and it has global variables, preventing it to be used in a multithreaded
context. CGAL can be completely parameterized through a template mechanism. However, by default, it internally uses
pointer-based data structures that do dynamic allocations, which has an impact of performance in a multithreaded context.

There are many references about constrained Delaunay triangulations, but most of them focus on its mathematical properties
and few of them focus on how to implement it.
A description of a reasonably efficient algorithm that works is given in one of the first references on this topic
\citep{sloan1992}. There are faster algorithms (divide and conquer, used in \cite{shewchuk96b}), but we will stick to
a simple algorithm to keep the implementation simple and easy to maintain. Moreover, since our coordinates are going to
be the result of intersections, execution time will be largely dominated by the predicates, so we can afford slightly
suboptimal combinatorics, provided that predicates invocation remains minimal. For that, we use a predicate cache,
as in the triangle-triangle intersection routine \secref{sec:tritri}.

Sloan's algorithm is reasonably easy
to implement, because it is (mostly) based on a single geometric operation: flip the edge common to two triangles.
Before we dive into the detail, let us see a high-level version of the algorithm:

$$
\begin{array}{ll}
  &\mbox{constrained\_Delaunay\_triangulation} \\
  &\mbox{\bf input:} \\
  &  \quad \mbox{a triangle } t_0 = (p_0, p_1, p_2)\\
  &  \quad \mbox{a list of vertices } p_i, i=3 \ldots N_v \mbox{ inside } t\\
  &  \quad \mbox{a list of edges } E_k=(i_k,j_k), k=1 \ldots N_e \\
  &\mbox{\bf output:} \\
  & \quad \mbox{the Delaunay triangulation of the } p_i \mbox{'s} \\
  & \quad \quad \mbox{constrained by the } E_k \mbox{'s and by } t\mbox{'s edges}. \\
  \hline
  \\
  (1) & \algfor i = 1 \algto N_v \\
  (2) & \quad \mbox{find the triangle } t \mbox{ that contains } p_i \\
  (3) & \quad \mbox{insert } p_i \mbox{ into } t \\
  (4) & \quad \mbox{push the three triangle edges opposite to } p_i \mbox{ onto } S \\
  (5) & \quad \mbox{Delaunize\_vertex\_neighbors}(p_i, S) \\
  (6) & \algend \\
  (7) & \algfor k = 1 \algto N_e \\
  (8) & \quad \mbox{enqueue the edges intersecting } (i_k, j_k) \mbox { onto } Q \\
  (9) & \quad N \leftarrow \mbox{constrain\_edges}(i_k, j_k, Q) \\
  (10) & \quad \mbox{Delaunize\_new\_edges}(N) \\
  (11) & \algend
\end{array}
$$

The algorithm starts from a single triangle $t_0 = (p_0,p_1,p_2)$ and inserts the vertices and the edges into it one by one.
It is made of two main blocks: \\

\paragraph*{\bf The first block} (lines (1) to (6)) inserts the vertices one by one in the triangulation, by first
locating the triangle $t$ that contains the point $p_i$. Then it splits this triangle into three (note that $p_i$ can be exactly
located on an edge, then the two triangles that share that edge are split into two, for a total of four new triangles). Then the
Delaunay condition is restored by the Delaunize\_vertex\_neighbors() function. This function recursively flips the edges
that violate the Delaunay condition and pushes the new triangle on the stack $S$ until the stack is empty. The reader is referred
to the original article \citep{sloan1992} for more details.

\paragraph*{\bf The second block} (lines (7) to (11)) inserts the constraints one by one in the triangulation. The first step (line (8)) detects
the edges that have an intersection with the constraint. This is done by ``walking the triangulation'' along the edge, one triangle
at a time, and testing for each triangle two edges (the third one is the one we came from). If one of the intersected edge is a
constraint, then this means we have detected a triple point, like in Figure \ref{fig:three_triangles}-C, where the intersection of edges
$e_1$ and $e_2$ generate a new vertex $I$. The intersected edges are pushed to a queue $Q$. Then the function constrain\_edges()
processes each ``flippable'' edge of the queue until the queue is empty. By ``flippable'', we mean that the two
triangles adjacent to the edge form a convex polygon. It can be proven that this process converges. Each time an edge is flipped,
the corresponding triangles are saved in a list $N$ of ``new'' triangles, finally processed by the Delaunize\_new\_edges() function
that flips edges until the Delaunay condition is satisfied everywhere. \\

My implementation is classical and follows this framework, with a couple of adaptations:
\begin{enumerate}
  \item first, we are going to compute a huge number of constrained
    Delaunay triangulations, in each individual triangle that has
    intersections. To keep performance acceptable, we are going to
    construct them in parallel. Dynamic memory allocation is a serious obstacle to efficient parallel code,
    because there is a global lock associated with the \verb|malloc()| function (or the \verb|new()| operator),
    so our data structure will be solely composed of \verb|std::vector|'s allocated once for
    all\footnote{except at the beginning when they will grow as needed, and later, they never shrink.}. We also need
    a data structure for the stack of triangles $S$ and the queue of triangles $Q$. I use a doubly connected
    list, represented as two additional \verb|std::vector|'s that store the forward and backward link;
  \item second, the algorithm manipulates edges. To keep things simpler, we systematically designate an edge
    through a triangle, and rotate the triangle in place in such a way that the designated edge is edge 0.
\end{enumerate}

  \begin{figure}[b]
    \centerline{
         \includegraphics[width=0.4\columnwidth]{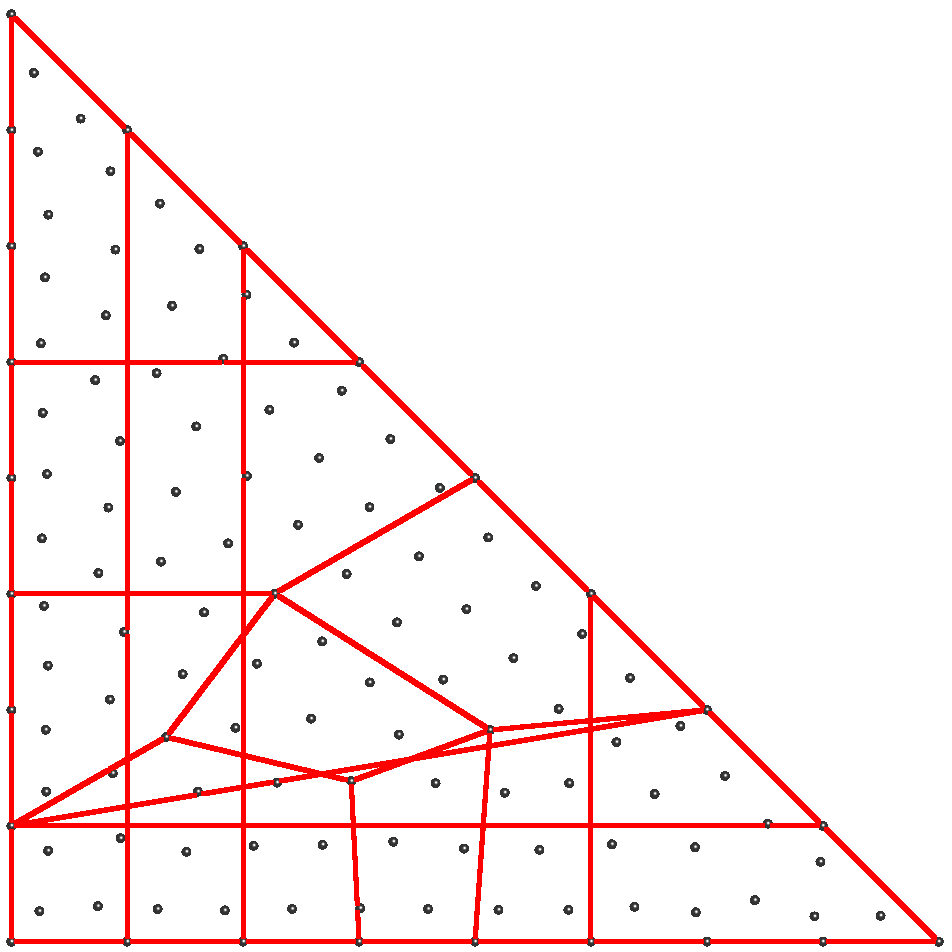}
         \hspace{5mm}
         \includegraphics[width=0.4\columnwidth]{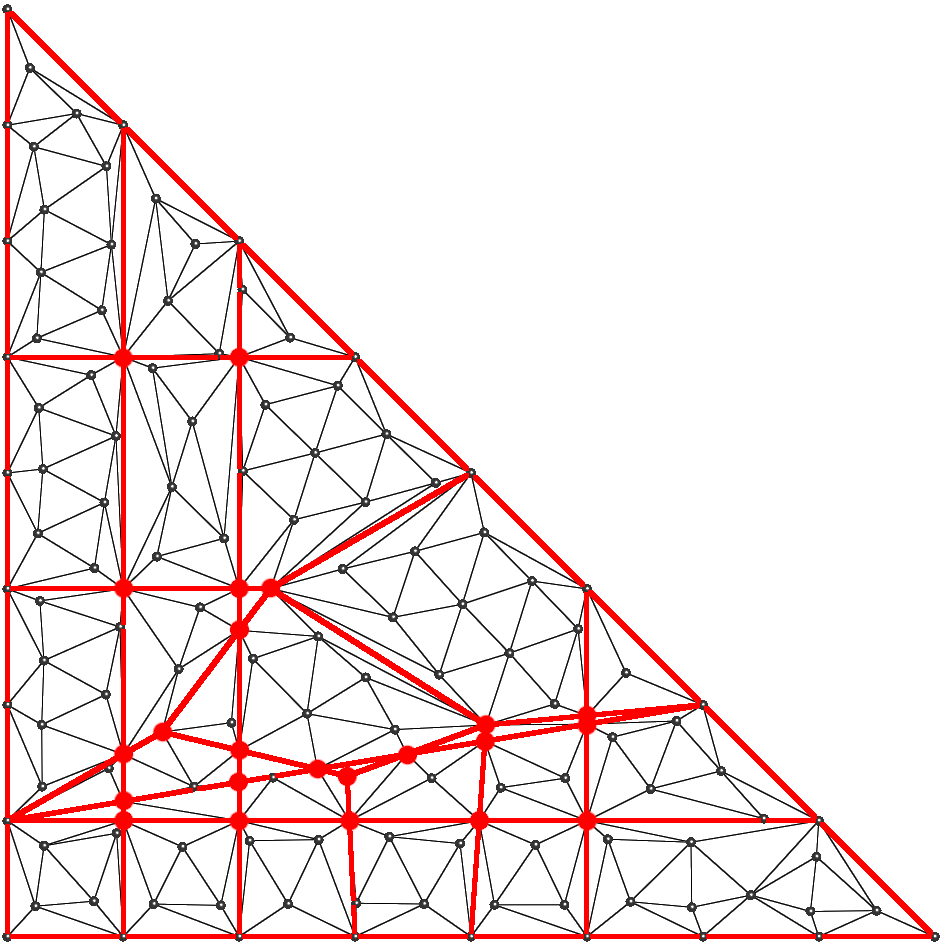}
    }
    \caption{Constrained Delaunay Triangulation in 2D. Left: input points and constraints.
      Note that some constraints have intersections. Right: the resulting triangulation. The vertices
      that correspond to constraint intersections were automatically inserted.}
    \label{fig:CDT}
  \end{figure}

  \begin{figure*}
    \centerline{\includegraphics[width=\textwidth]{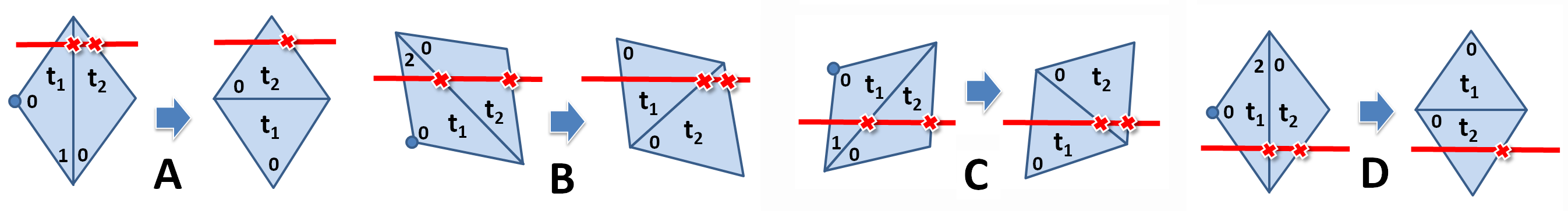}}
    \caption{ The four configurations of triangle flips that can be encountered during the constraint enforcement phase. The constrained edge $(i,j)$ is shown in red. The four configurations are determined by (1) the orientation of $t_1.v_0$, shown as a (pale) blue dot, with respect to $(i,j)$ and (2) whether $t_2.v_0$ corresponds to $t_1.v_1$ or $t_1.v_2$. By exploiting the combinatorial information, one tests whether a triangle's edge intersects the constraint $(i,j)$ with a single ${\tt orient\_2d}$ invocation (instead of up to 4)}
    \label{fig:flips}
  \end{figure*}

  The algorithm to restore the Delaunay condition around new vertices and around new edges is as in the classical
  implementations of the Bowyer-Watson algorithm. The algorithm to constrain the edges deserves more details, because
  the way the edges are systematically manipulated through triangles reveals an optimization that avoids most
  invocations to the ${\tt orient\_2d}$ predicate. The classical version is as follows:

$$
\begin{array}{ll}
  &\mbox{constrain\_edges} \\
  &\mbox{\bf input:} \\
  &  \quad \mbox{the edge } i,j \mbox{ to be constrained}\\
  &  \quad \mbox{a queue } Q \mbox{ initialized with the edges intersected by } (i,j)\\
  &\mbox{\bf output:} \\
  & \quad \mbox{the set } N \mbox{ of new edges} \\
  \hline
  \\
  (1) & \algwhile Q \mbox{ is not empty } \\
  (2) & \quad (v_1, v_2) \leftarrow \mbox{dequeue}(Q) \\
  (3) & \quad \algif \mbox{the two triangles incident to } (v_1,v_2) \mbox{ form a convex quad } \\
  (4) & \quad \quad \mbox{swap\_edge}(v_1,v_2) \\
  (5) & \quad \quad \algif (v_1, v_2) \cap (i,j) \neq \emptyset \algthen \\
  (6) & \quad \quad \quad \mbox{enqueue}(Q,(i,j)) \\
  (7) & \quad \quad \algelse \\
  (8) & \quad \quad \quad N \leftarrow N \cup { (v_1,v_2) } \\
  (9) & \quad \quad \algend \\
  (10) & \quad \algelse \\
  (11) & \quad \quad \mbox{enqueue}(Q,(i,j)) \\
  (12) & \quad \algend \\
  (13) & \algend
\end{array}
$$

\new{I make here a simple yet important observation that spares 3/4 of the calls to the} \verb|orient_2d| \new{predicate}: The segment-segment intersection test in line (5) that normally requires up to 4 invocations of the ${\tt orient\_2d}$ predicate
can be replaced with a single invocation and some combinatorics. Remember that each edge $(v_1, v_2)$ is systematically
manipulated through a triangle $(v_0, v_1, v_2)$, rotated in such a way that its vertex $v_0$ is opposite to the edge
under consideration. Then, as shown in Figure \ref{fig:flips}, there are four different configurations for the
pair of triangles $t_1=(v_0, v_1, v_2)$ and $t_2$ adjacent to $t_1$ along $(v_1,v_2)$, depending on whether $t_2$ is in the queue $Q$
of intersected edges\footnote{that is tested in O(1) using a per-triangle array of marks.}, depending on ${\tt orient\_2d(i,j,v_0)}$ and depending on whether $t_2$'s first vertex is $v_1$ or $v_2$:
\begin{itemize}
\item $t_2 \notin Q$ then after swapping there can't be any intersection \\then $(v_1,v_2)$ is pushed to $N$ (new edge);
\item config. A: {(${\tt orient \_2d}(v_0,i,j) < 0$ and $t2.v0 = t1.v1$)} \\then $t_1$ is pushed to $N$ (new edge);
\item config. B: {(${\tt orient \_2d}(v_0,i,j) < 0$ and $t2.v0 = t1.v2$)} \\then $t_1$ is pushed to $Q$ (still has an intersection);
\item config. C: {(${\tt orient \_2d}(v_0,i,j) > 0$ and $t2.v0 = t1.v1$)} \\then $t_1$ is pushed to $Q$ (still has an intersection);
\item config. D: {(${\tt orient \_2d}(v_0,i,j) > 0$ and $t2.v0 = t1.v2$)} \\then $t_1$ is pushed to $N$ (new edge)
\end{itemize}

I give also here a short comment on another part of the algorithm, that determines the list of edges intersected by
the constraint $(i,j)$. This part of the algorithm is conceptually simple (just walk along the triangles according
to ${\tt orient\_2d()}$, very similarly to what is done for locating a point), but there are three subtelties:
\begin{itemize}
\item whenever an existing vertex $v$ lands exactly on the constrained edge $(i,j)$, one needs to traverse the fan of
  triangles incident to $v$ in order to find the next triangle;
\item there can be co-linear overlapping constraints, hence one needs to store in each
  segment the list of constraints it belongs to. In my implementation they are chained.
  This information is required for instance when classifying the triangles;
\item whenever an intersecting edge is also a constraint, one needs to insert the intersection of both constraints into the
  triangulation, and delaunay-ize its neighborhood, before enforcing the edge constraints.
\end{itemize}

An example of a 2D constrained Delaunay triangulation with intersecting constraints is shown in Figure \ref{fig:CDT}.
The constrained segments are shown in red. Some of them have intersections (red dots).
The algorithm automatically detects the intersections, inserts them in the triangulation, respects all the constraints
and the Delaunay criterion. \\

Let us see now how the geometric part of the algorithm works. First, one can notice that the constrained Delaunay triangulation
is mostly combinatorial. The only places where geometrical information is used is a set of three functions. The first two ones
are \emph{predicates}, that return a sign (negative, zero or positive):
\begin{itemize}
   \item \verb|orient(i,j,k)|: computes the 2D orientation of the triangle with vertices $i$, $j$ and $k$. It is used in many places,
     to detect the triangle that contains an inserted point, to test for intersecting edges, and to test whether an edge is flippable.
   \item \verb|incircle(i,j,k,l)|: this corresponds to the Delaunay condition (the circumscribed circle of each triangle should not
     contain any vertex). This function is symbolically perturbed, in such a way that it never returns zero
     (see \cite{DBLP:journals/corr/EdelsbrunnerM94a}). This ensures a unique triangulation for configurations with cocyclic vertices.
\end{itemize}
The third function is a \emph{construction}, that creates new geometry:
\begin{itemize}
  \item \verb|create_intersection(i,j,k,l)|: create a new vertex that corresponds to the intersection between edges $(i,j)$ and $(k,l)$
\end{itemize}

So now the question is how to implement these three functions given that some of our vertices are given as intersections between
segments and triangles. The strategy here is to keep the algorithm as near as possible to ``simply implement
the math''. In other words, it means we are pushing most of the difficulties towards the geometric kernel (predicates and constructions)
to keep the overall structure of the algorithm simple (nearly a verbatim copy of the textbook algorithm). So
we are going to compute the coordinates of the intersection points explicitely. Since these coordinates are rational fractions, and
since computations for $x,y$ and $z$ are inter-related, it is reasonable to represent the intersection points in homogeneous coordinates
$\hat{\bp} = [ x\ y\ z\ w ]$ that corresponds to a 3D point $\bp = [ x/w\ y/w\ z/w ]$. Each individual $x,y,z,w$ coordinate is represented
in an \emph{exact number type} that exactly implements addition, subtraction and multiplication. \new{This is a natural representation for intersections, see \cite{DBLP:journals/corr/EdelsbrunnerM94a} for an excellent tutorial on the topic, see also CGAL Homogeneous Kernel, or recent works such as \cite{NehringWirxel2021FastEB}.} We shall see later
\secref{sec:arithmetics} two alternatives to implement exact number types, and how to implement all the predicates that we need. \\

The input of the constrained Delaunay triangulation is a triangle $t_1$ and the list of segments to be inserted into $t_1$. Each segment's
extremity is encoded symbolically, as a triple $\sigma_1, t_2, \sigma_2$, indicating the location of the intersection within $t_1$,
the other facet $t_2$ and the location of the intersection within $t_2$. Note that there can be also individual points (for instance, when a
vertex is exactly located in a facet). The first task to do is computing the coordinates of each intersection. Depending on the nature
of $\sigma_1$ and $\sigma_2$, there are three different cases, plus an additional case for intersecting constraints \new{(four cases in total)}:

\begin{itemize}
  \item{initial vertex:} just convert the input point (with floating-point coordinate) to the arbitrary precision representation;
  \item{edge $e_1$ $\cap$ triangle $t_2$ in 3D or edge $e_1$ $\cap$ edge $e_2$ in 3D};
  \item{edge $\cap$ edge in 2D};
  \item{intersection of two constraints}
\end{itemize}

  Let us detail now how to compute the coordinates of the intersection point $\bI$ for the last three configurations:

  \paragraph*{\bf edge $e_1$ $\cap$ triangle $t_2$ in 3D or edge $e_1$ $\cap$ edge $e_2$ in 3D:}
    Let $\bq_1$ and $\bq_2$ denote the extremities of the edge and let $\bp_1$,$\bp_2$ and $\bp_3$ denote the vertices of the triangle.
    We are in 3D if $\bq_1$ and $\bq_2$ are not both in the supporting plane of $\bp_1$,$\bp_2,\bp_3$ (in other words, at least one
    of \verb|orient_3d|$(\bp_1,\bp_2,\bp_3,\bq_1)$ and \verb|orient_3d|$(\bp_1,\bp_2,\bp_3,\bq_2)$  is non-zero). Note that if we are
    in 3D, the \emph{edge} $e_1$ $\cap$ \emph{edge} $e_2$ configuration can be seen as a particular case of the
    \emph{edge} $e_1$ $\cap$ \emph{triangle} $t_2$ case, by replacing $e_2$ by the triangle $t_2$ it comes from. At the previous
    step, we already determined that the intersection exists, so to compute its coordinates we just need to compute a line-plane
    intersection, without needing to check that the intersection is in the segment and in the triangle. The intersection point
    $\bI$ belongs to the line:
$$
    \bI = \bq_1 + t (\bq_2 - \bq_1) \quad t \in \R
$$
    and to the plane:
$$
    (\bI - \bp_1) \cdot \bN = 0 \quad \mbox{where } \bN = (\bp_2 - \bp_1) \times (\bp_3 - \bp_1).
$$
By substitution, one gets:
$$
\begin{array}{l}
  t = \frac{(\bp_1 - \bq_1) \cdot \bN}{(\bq_2 - \bq_1) \cdot \bN} \quad ; \quad \bI = \mix(t, \bq_1, \bq_2) \\[3mm]
  \mbox{where } \mix(t, \bq_1, \bq_2, t) = \bq_1 + t (\bq_2 - \bq_1) = (1-t) \bq_1 + t \bq_2.
\end{array}
$$

One needs to keep in mind that these computations are made with exact numbers (more on this in \secref{sec:arithmetics}).
Since our exact numbers only support addition, subtraction and product, and since $t$ is a rational number,
the intersection $\bI$ will be represented in homogeneous coordinates. So we need an implementation of $\mix(t,\bq_1,\bq_2)$
that takes two points $\bq_1, \bq_2$ with floating-point coordinates, an exact rational parameter $t$, and that returns
a point with homogeneous coordinates:

  $$
  \mbox{mix}\left(\frac{a}{b}, \bq_1, \bq_2, \right) = \frac{a}{b} \bq_2 + \frac{b-a}{b} \bq_1 =
    \left[
      \begin{array}{c}
        a \bq_2 + (b-a) \bq_1 \\
        b
      \end{array}
    \right]_h
  $$
where the $h$ subscript indicates that the point has homogeneous coordinates. Note that one could also use 3d vectors with
the $x$, $y$, $z$ coordinates as independent rational numbers instead of homogeneous coordinates.

\paragraph*{\bf edge $\cap$ edge in 2D:} computing the intersection of two coplanar edges
  cannot be done by the formula above (because for coplanar edge and triangle, the denominator is zero),
  so we compute the intersection in 2D (using the $2\times2$ Cramer formula),
  and lift it to 3D using the 3D points $P_1$ and $P_2$ associated with $p_1$ and $p_2$:
  $$
  \begin{array}{lcl}
    \bI & = & \mbox{mix}(t,\bp_1, \bp_2) \quad
     \mbox{where:} \\[2mm]
    t & = & \mbox{det}(\bq'_1-\bp'_1, \bq'_2-\bq'_1) \ /\ \mbox{det}(\bp'_2-\bp'_1,\bq'_2-\bq'_1),
  \end{array}
  $$
  and where $\bp_1, \bp_2, \bq_1, \bq_2$ denote the (3D) extremities of the two segments, and
  $\bp'_1, \bp'_2, \bq'_1, \bq'_2$ denote the (2D) projected extremities of the two segments.

\paragraph*{\bf intersection of two constraints:} the last possible configuration for a constructed
intersection point is encountered whenever two constrained segments have an intersection (see
Figure \ref{fig:three_triangles}). Clearly, it is possible to reuse the 2D segment intersection
formula above, and lifting it to 3D (by $t$-mixing the 3D points instead of the 2D points). However,
the resulting expression has two nested levels of exact operations (expression of $t$ and $\mix()$).
It especially has an impact on performance when using the arithmetic expansions \secref{sec:expansions}.
By recalling that an intersection between two constraints systematically corresponds to an intersection
between three triangles, one can obtain a simpler and more symmetric equation for the intersection.
The intersection between three triangles can be obtained easily, using the $3 \times 3$ Cramer formula:

$$
\begin{tiny}
  \left[
  \begin{array}{ccc}
  a_{11} & a_{12} & a_{13} \\
  a_{21} & a_{22} & a_{23} \\
  a_{31} & a_{32} & a_{33} \\
  \end{array}
  \right]
  \left[
  \begin{array}{c}
      x_1 \\ x_2 \\ x_3
  \end{array}
  \right] =
  \left[
  \begin{array}{c}
      y_1 \\ y_2 \\ y_3
  \end{array}
  \right] \quad \Rightarrow
\end{tiny}
  $$

$$
  \begin{tiny}
  \begin{array}{lcl}
    x_1 & = &
  \left|
  \begin{array}{ccc}
  y_{1} & a_{12} & a_{13} \\
  y_{2} & a_{22} & a_{23} \\
  y_{3} & a_{32} & a_{33} \\
  \end{array}
  \right| /
  \left|
  \begin{array}{ccc}
  a_{11} & a_{12} & a_{13} \\
  a_{21} & a_{22} & a_{23} \\
  a_{31} & a_{32} & a_{33} \\
  \end{array}
  \right| \\ \vspace{1mm} \\
    x_2 & = &
  \left|
  \begin{array}{ccc}
  a_{11} & y_{1} & a_{13} \\
  a_{21} & y_{2} & a_{23} \\
  a_{31} & y_{3} & a_{33} \\
  \end{array}
  \right| /
  \left|
  \begin{array}{ccc}
  a_{11} & a_{12} & a_{13} \\
  a_{21} & a_{22} & a_{23} \\
  a_{31} & a_{32} & a_{33} \\
  \end{array}
  \right| \\ \vspace{1mm} \\
    x_3 & = &
  \left|
  \begin{array}{ccc}
  a_{11} & a_{12} & y_{1} \\
  a_{21} & a_{22} & y_{2} \\
  a_{31} & a_{32} & y_{3} \\
  \end{array}
  \right| /
  \left|
  \begin{array}{ccc}
  a_{11} & a_{12} & a_{13} \\
  a_{21} & a_{22} & a_{23} \\
  a_{31} & a_{32} & a_{33} \\
  \end{array}
  \right|
  \end{array}
  \end{tiny}
  $$

\begin{figure}
  \includegraphics[width=0.4\columnwidth]{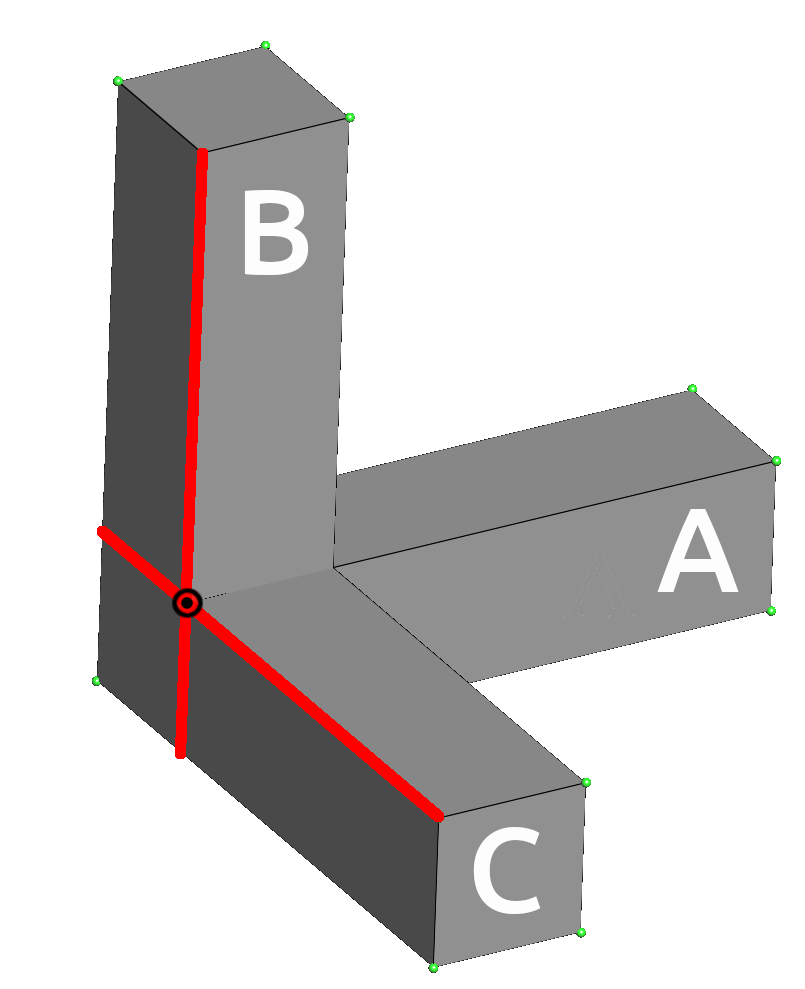}
  \caption{The intersection between two edges of B and C (in red) landing exactly on a corner of A (highlighted).}
  \label{fig:three_rods}
\end{figure}

Given our three triangles $(\bp_1, \bp_2, \bp_3)$, $(\bq_1, \bq_2, \bq_3)$ and $(\br_1, \br_2, \br_3)$,
the intersection $\bI$ is given by:
$$
\begin{tiny}
\bI = \left[\begin{array}{l}
    \left|
    \begin{array}{ccc}
      B_x & N1_y & N1_z \\
      B_y & N2_y & N2_z \\
      B_z & N3_y & N3_z
    \end{array}
    \right|\\ \vspace{1mm}\\

    \left|
    \begin{array}{ccc}
      N1_x & B_x & N1_z \\
      N2_x & B_y & N2_z \\
      N3_x & B_z & N3_z
    \end{array}
    \right| \\ \vspace{1mm}\\

    \left|
    \begin{array}{ccc}
      N1_x & N1_y & B_x \\
      N2_x & N2_y & B_y \\
      N3_x & N3_y & B_z
    \end{array}
    \right| \\\vspace{1mm}\\
     \left|
    \begin{array}{ccc}
      N1_x & N1_y & N1_z \\
      N2_x & N2_y & N2_z \\
      N3_x & N3_y & N3_z
      \end{array}
      \right|
  \end{array}\right]_h
  \end{tiny}
$$
where $\bN_1, \bN_2, \bN_3$ denote the normal vectors of the three triangles, and where
$\bB = [ \bN_1 \cdot p_1,  \bN_2 \cdot q_1, \bN_3 \cdot r_1]$. As in the previous case,
one does not need to check whether the intersection of the supporting planes belongs to the triangles.
We already know it does because it is the triangle-triangle intersection function that generated the constraints.

  \begin{figure*}
    \centerline{\includegraphics[width=\textwidth]{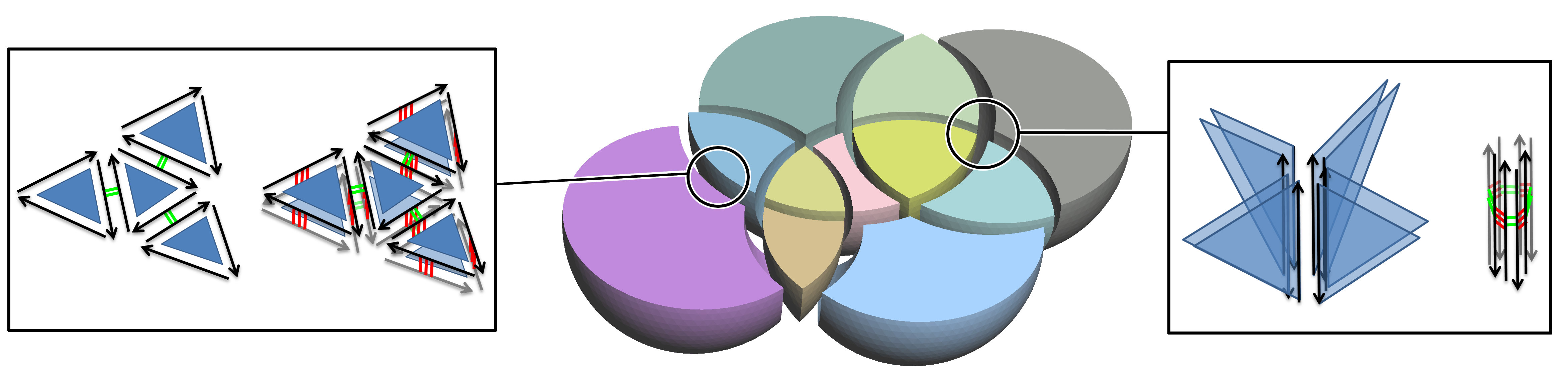}}
    \caption{Weiler model and 3-Maps. A 3-map is composed of a set of darts (black arrows), a permutation $\sigma_1$ that
      connects each dart to its successor around a triangle, an involution $\alpha_2$ (double green lines) that connects
      each dart to the opposite dart in a neighboring triangle and an involution $\alpha_3$ (triple red lines) that
      connects each dart to the opposite dart in the twin sheet. Non-manifold edges form bundles of more than four darts,
      and need a geometric radial sort.}
    \label{fig:weiler}
  \end{figure*}

At this point, we are able to generate the coordinates of the constructed points for the three possible configurations
for an intersection (edge-triangle or edge-edge in 3D, edge-edge in 2D, and three triangles). The combinatorial part
of the constrained Delaunay triangulation queries the geometry through the classical $\tt orient2d$ and $\tt incircle$
predicates. We shall see later how to implement them, as well as the other ones used by the subsequent steps of
the algorithm \secref{sec:arithmetics}.

\paragraph*{The global vertex table}
\label{sect:global_vertex_table}

We have computed the exact coordinates of the intersection points, and we have inserted all the intersection
segments into the triangles.
Consider an intersection $\bI$ between a triangle $t$ and an edge $e$ shared by two triangles $t_1$ and $t_2$. The
intersection $\bI$ will be generated twice (once when triangulating $t_1$ and once when triangulating $t_2$).
Clearly, one could use a key made of integer ids that define $e$ and $t$ to index a global vertex table. However, some
nasty configurations may appear. There are two cases:
\begin{itemize}
  \item two different intersections may land exactly on the same point;
  \item an intersection may land exactly on a vertex that exists in the input meshes.
\end{itemize}
Again, considering these cases is not academic paranoia taking the exact computation paradigm too seriously,
they often occur in practice. Figure \ref{fig:three_rods} shows one of the simplest example, with three intersecting
rods forming a corner. The point in red is the intersection between the two highlighted edges of rods B and C. It exactly
corresponds to a vertex of A. This type of configuration is very likely to appear in CAD objects generated by CSG. \\

In the exact geometry paradigm, in a certain sense, ``geometry is combinatorics'', hence,
to handle all these configuration, the idea is to use a global vertex table indexed by the coordinates of the points.
Implementing a table (e.,g., with \verb|std::map| from the C++ STL)
requires a function to compare two keys (two points in our case), so we are going to use the lexicographic
order on the point's coordinates. However,
one needs to remember that the representation of a point in homogeneous coordinates is non-unique,
hence one cannot simply use the lexicographic order on $x,y,z,w$. Instead of that, we compare the (rational) Euclidean
coordinates of the points $x/w, y/w, z/w$, which can be done as follows with our exact number type that does not have
division:

$$
\sign\left(\frac{x_1}{w_1} - \frac{x_2}{w_2}\right) = \sign(w_1) \times \sign(w_2) \times \sign(w_2 x_1 - w_1 x_2)
$$

We shall see \secref{sec:arithmetics} some optimizations that can be made
depending on the used exact arithmetic kernel. \\

To keep the size of the table reasonable, the vertices of the input meshes are not inserted into the table: they are instead
processed ``the other way round'', at the end of the algorithm,
a post-processing phase queries the table with all the input point and merges the co-located ones.

\subsection{Mesh boolean operations and CSG}

\subsubsection{The Weiler model}
\label{sec:Weiler}

To evaluate boolean expressions, we construct a combinatorial representation of the relations within a volumetric
mesh, called the Weiler model \citep{DBLP:journals/cga/Weiler85}. In what follows, I shall use the notations
of combinatorial maps \citep{DBLP:conf/stacs/Lienhardt88}. \new{This notion was introduced for writing the algebraic
specification of geometric algorithm operating on meshes. In our case it is useful as a way of compacly
writing both the algorithmic description of the method and its computer implementation.
In a combinatorial map, a mesh is represented as a set of objects (called \emph{darts})
and connections between them. Combinatorial maps come also with higher level operations to
manipulate and to navigate in portions of the mesh.
Here we use a 3-dimensional combinatorial map (3-map) to represent the Weiler model (or 3d mesh arrangement). Intuitively,
in a 3-map, the combinatorial elements (darts) are very similar to the classical halfedges used in surface meshes, with
additional volumetric links \new{that connect the boundaries of adjacent volumes (Fig. \ref{fig:weiler}). Zones where
  two volumes touch each other are represented as ``twin sheets'' of darts with opposite orientations connected
with volumetric links.} \\
}

\new{Formally,} a 3-dimensional combinatorial map (3-map) is defined as a quadruple $(\D, \sigma_1, \alpha_2, \alpha_3)$,
where $\D$ is a set of $N$ discrete elements called \emph{darts}, symbolized as black arrows in
Figure \ref{fig:weiler}, that can be identified with the integers $[1..N]$. The three functions
$\sigma_1, \alpha_2, \alpha_3$, acting on the set of darts $\D$, are defined as follows:
\begin{itemize}
\item $\sigma_1$ is a permutation, that maps each dart to its successor around a triangle;
\item $\alpha_2$ is an involution ($\alpha_2 \circ \alpha_2 = \mbox{Id}$), that maps each dart to the opposite dart in the
  neighboring triangle within the same surface (double green lines in Figure \ref{fig:weiler};
\item $\alpha_3$ is an involution that maps each dart to the opposite dart within the ``twin'' surface sheet (triple red lines in Figure \ref{fig:weiler}).
\end{itemize}

As can be seen, each triangle is composed of three darts. In practice,
only triangles and $\alpha_2, \alpha_3$ are stored explicitly. A triangle
with index $t$ corresponds to three darts $3t$, $3t+1$, $3t+2$, and
the permutation $\sigma_1$ is given by $\sigma_1 = d - (d \mod 3) + (d+1) \mod 3$.

It is worth mentioning that $\alpha_2$ and $\alpha_3$ systematically connect
darts with opposite orientations. As a consequence, non-orientable surfaces
(Moebius strip, klein bottle \ldots) cannot be represented with a 3-map. For the interested
reader, there exists a notion of generalized map \citep{DBLP:journals/ijcga/Lienhardt94} that represents
a wider class of objects (cellular quasi-manifold), comprising non-orientable surfaces.
In our context, since we compute CSG and boolean operations, the surfaces we consider
are supposed to define closed volumes, hence they are orientable.

One can also define the \emph{orbit} $<\beta_1, \beta_2, \ldots \beta_n>(d)$ as the
set of darts that can be recursively reached by traversing all links
$\beta_1, \beta_2, \ldots \beta_n$ from a dart $d$. There are 8 possible types of orbits, the
following ones are of particular interest:
\begin{itemize}
   \item {\bf individual triangle}: $<\sigma_1>(d)$
   \item {\bf shell}: $<\sigma_1,\alpha_2>(d)$, that is, the boundary of one of the
             colored volumetric regions in Figure \ref{fig:weiler};
   \item {\bf bundle}: $<\alpha_2,\alpha_3>(d)$, that correspond to the fan of triangles incident to the
     same edge. A bundle is non-manifold if it has more than four darts (Figure \ref{fig:weiler}-right).
     Bundles are referred to as ``radial edges'' in Weiler's parlance;
   \item {\bf connected component}: $<\sigma_1, \alpha_2,\alpha_3>(d)$.
\end{itemize}

In addition, we define a notion of patch. The patch incident to a dart $d$ is
defined by $< \sigma_1,\bar{\alpha_2}>(d)$, where $\bar{\alpha_2}(d)$ is defined by:
$$
\begin{array}{lcl}
  \bar{\alpha_2}(d) & = & \alpha_2(d) \mbox{ if the bundle incident to } d \mbox { has 4 darts } \\
\bar{\alpha_2}(d) & = & d \mbox{ otherwise. }
\end{array}
$$
Note that the zone where two shells are in contact corresponds to two different patches, one for each side.

\subsubsection{Constructing the Weiler model}

Now our goal is to construct the Weiler model from the output of the co-refinement phase. We start by duplicating all
the triangles, and connecting each dart to its counterpart with $\alpha_3$ links. Then we identify the bundles by
sorting all the darts in lexicographic order based on the indices of their two extremities. In the sorted list of darts,
the bundles are easy to find as contiguous sequences with the same extremities. For each bundle, they are two cases to
consider:

\begin{itemize}
\item {\bf the bundle has 4 darts:} this is the easy case, that corresponds to a manifold edges. We just need to
     create two $\alpha_2$ links connecting each part of darts;
   \item {\bf the bundle has more than 4 darts:} the bundle corresponds to a non-manifold edge (like in Figure \ref{fig:weiler}-right). To determine which darts should be connected with $\alpha_2$ links, one needs to (geometrically) sort the triangles around
     the non-manifold edge, an operation referred to as \emph{radial sort} in \cite{DBLP:journals/cga/Weiler85}.

     To define a total order of the darts arround a halfedge, one picks one of the darts
     $h_0$ as the origin, and one uses two predicates:

     \begin{itemize}
     \item ${\tt orient}(h_1,h_2) = {\tt orient\_3d}(\bp_1,\bp_2,\bp_3,\bp_4)$ where
       $\bp_1$ and $\bp_2$ are the two extremities of the radial edge and $\bp_3$ and
       $\bp_4$ the two vertices opposite to the radial edge in the two triangles incident
       to $h_1$ and $h_2$;
     \item ${\tt Norient}(h_1,h_2) = \mbox{sign}(\bn_1 \cdot \bn_2)$ where $\bn_1$ and $\bn_2$ denote
       the normals to the triangles incident to $h_1$ and $h_2$.
     \end{itemize}

     For a given dart $h$ in the bundle, the two signs given by
     ${\tt orient}(h_0,h)$ and ${\tt Norient}(h_0,h)$ dermine four quadrants around the
     radial edge. If the two darts $h_1,h_2$ to be compared are in a different quadrant,
     then their relative order is known. If $h_1,h_2$ are in the same quadrant,
     then their relative order is determined by ${\tt orient}(h_1,h_2)$.
\end{itemize}

As for the predicates used in the other stages of the pipeline, these
two predicates are filtered using interval arithmetics. The vector
$\bp_2 - \bp_1$ of the radial edge and the normal vector $\bn_0$ are computed
at the beginning of radial sort, in both interval and exact arithmetic.

\begin{wrapfigure}{l}{0.19\textwidth}
  \includegraphics[width=0.2\textwidth]{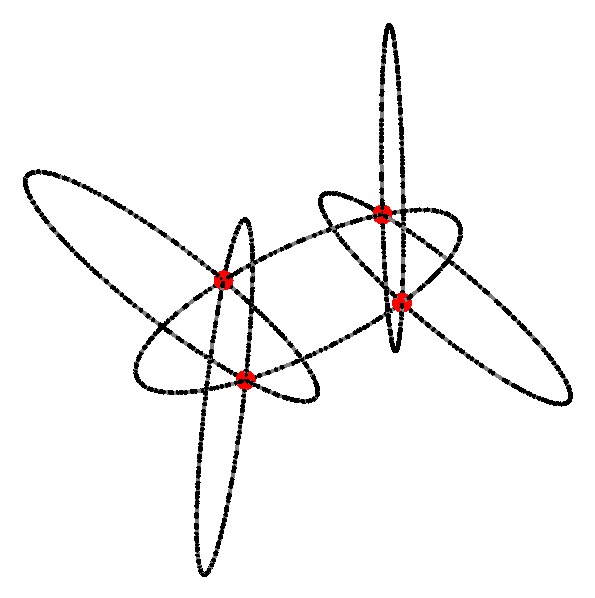}
\end{wrapfigure}
\noindent Note that in these predicates, the two points $\bp_1$ and $\bp_2$
along the radial edge are intersection points, hence they are
represented with homogeneous exact coordinates. As a consequence, even
with the interval filters, sorting all the radial edges takes a
significant amount of time. However, one can gain significant performance by noticing
that the order around a radial edge
remains the same when traversing polygonal lines incident to the same set of patches.
Hence one can propagate the order along radial polylines, stopping at the vertices
incident to more than two bundles, shown in red in the small figure.

Once all the radial edges are sorted, the combinatorial links
$\sigma_0, \alpha_2$ and $\alpha_3$ can be set:
\begin{itemize}
\item The links $\sigma_0$ (that connect each dart to its successor around the triangle)
  and $\alpha_3$ (that connect each part of triangles adjacent to the same patch) are
  trivial to obtain (they do not depend on the radial sort);
\item around each bundle $d_1, d_2 \ldots d_k$, an $\alpha_2$ connection is created
  between each pair of darts $\alpha_3(d_i)$ and $d_{i \oplus 1}$, where $i \oplus 1$
  denotes $i + 1$ modulo $k$.
\end{itemize}

\subsubsection{Classifying}
\label{sec:classify}

Once the Weiler model is constructed, the next step consists in \emph{classifying}
the triangles (step 3 in the processing pipeline shown in Figure \ref{fig:CSG_overview}
on page \pageref{fig:CSG_overview}). \new{The method is very similar to what is done in
  \cite{10.1145/2897824.2925901} (with the difference that the interface between two
  volumetric regions is represented by two sheets instead of a single sheet, making
  classification a little bit easier). }

The input of the classification phase is:
\begin{itemize}
\item at the very beginning of the pipeline, the input facets
  correspond to a set of $N$ operands $\OO_1, \OO_2, \ldots \OO_n$. Each operand
  is represented by its (closed) boundary. The information of which facet comes from
  which operand is represented by sets $B(t)$ associated to each input triangular
  facet $t$:
$$
 B(t) = \{ \OO_i | t \subset \partial \OO_i \}.
$$
 In practice, $B(t)$ can be represented by a bitvector associated with $t$, and the
 $i$-th bit indicates whether $t$ belongs to $\partial \OO_i$.
 Note that the same input triangle may belong to the boundaries of several operands,
 for instance whenever two operands touch along a common surface. During the co-refinement
 phase, whenever two identical triangles $t_1$ and $t_2$ are merged into a single
 output triangle $t$, the associated sets are also merged: $B(t) = B(t_1) \cup B(t_2)$
 (at the implementation level, in terms of bitvectors, they are ORed);
\item a boolean expression $E(b_1, b_2, \ldots b_N)$ that takes as an
  argument a vector of $N$ booleans and that returns a boolean. For
  instance, for the union of two operands, this expression corresponds
  to $(b_1 \mbox{ or } b_2)$. For the intersection, it corresponds to
  $(b_1 \mbox{ and } b_2)$. For the difference, it corresponds to $(b_1
  \mbox{ and not } b_2)$.  It can have an arbitrary number of
  operands. For instance, in the example shown in Figure
  \ref{fig:CSG_overview} P. \pageref{fig:CSG_overview}, the
  expression is $E(b_1,b_2,b_3,b_4) = ((b_1 \mbox{ or } b_2 \mbox{ or } b_3) \mbox{
    and not } b_4)$.
\end{itemize}

The classification phase aims at finding all the darts that are on the boundary
of the object $\OO_E$ defined by the boolean expression $E$. It is done in two phases:
\begin{itemize}
\item For each dart $d$, determine the set of objects $I(d) = \{ \OO_i | d \in \OO_i \}$.
  The algorithm to compute the $I(d)$'s will be explained later. We consider
  that a dart $d$ belongs to an object $\OO_i$ if the triangle $t(d)$ it is incident to
  is included in $\OO_i$ or if $t(d)$ is included in the boundary $\partial \OO_i$ and has
  a normal vector that points inwards $\OO_i$. In other words, considering the two charts
  that cover the boundary of $\OO_i$ (connected with $\alpha_3$ links,
  see Figure \ref{fig:weiler}-left), one of them is considered to be outside $\OO_i$ and
  the other one inside $\OO_i$. Put differently, whenever one crosses an $\alpha_3$ link
  from one of these darts, one moves from inside to outside or from outside to inside;
\item once the $I(d)'s$ are computed, one can characterize the darts on
  $\partial \OO_E$ as follows:
$$
  d \in \partial \OO_E \Leftrightarrow
     \mbox{ not } E(I(d)) \mbox{ and }  E(I(\alpha_3(d))),
$$
in other words, $d$ is on the (external) boundary of $\OO_E$ if $d$ is outside $\OO_E$
and if one gets inside $\OO_E$ by traversing the $\alpha_3$ link from $d$.
\end{itemize}

Let us see now how to compute the sets $I(d)$.
We first consider how to classify all the darts in a single connected component, starting
from a dart $d$, and knowing the set of objects $I(d)$ that contain $d$.
Consider for now that $d$ is on the external boundary of the connected component,
hence $I(d) = \emptyset$.

$$
\begin{array}{ll}
  &\mbox{classify\_component(d,B)} \\
  &\mbox{\bf input:}
     \quad \mbox{a dart } d
      \mbox{ and the set } I(d) = \{ \OO_i\ |\ d \in  \OO_i \} \\
  &\mbox{\bf output:}
    \quad \mbox{the sets }
      I(d^\prime) \mbox{ for } d^\prime \in <\sigma_0, \alpha_2, \alpha_3>(d) \\
  \hline
  \\
  (1)  & \mbox{push}(S,d); \mbox{mark}(d) \\
  (2)  & \algwhile S \mbox{ is not empty} \\
  (3)  & \quad d_1 \leftarrow \mbox{pop}(S) \\
  (4)  & \quad \algfor d_2 \in \{ \sigma_1(d_1), \alpha_2(d_1) \} \\
  (5)  & \quad \quad \algif d_2 \mbox{ is not marked } \algthen \\
  (6)  & \quad \quad \quad I(d_2) \leftarrow I(d_1) \quad ; \quad \mbox{push}(S,d_2) \quad ; \quad \mbox{mark}(d_2) \\
  (7)  & \quad \quad \algend \\
  (8)  & \quad \algend \\
  (9)  & \quad d_3 \leftarrow \alpha_3(d_1) \quad ; \quad t \leftarrow t(d_1) \\
  (10) & \quad \algif d_3 \mbox{ is not marked } \algthen \\
  (11) & \quad \quad I(d_3) \leftarrow
     \left( I(d_1) \cap C_{B(t)} \right)  \cup
     \left( C_{I(d_1)} \cap B(t) \right) \\
  (12) & \quad \quad \mbox{push}(S,d_3) \quad ; \quad \mbox{mark}(d_3) \\
  (13) & \quad \algend \\
  (14) & \algend
\end{array}
$$

The algorithm greedily traverses all the $\sigma_1, \alpha_2$ and $\alpha_3$ links from $d$.
In steps (4-8), $\sigma_1$ and $\alpha_2$ links are traversed, one stays within the same
shell, hence $I(d)$ is propagated.
In step (11), an $\alpha_3$ link is traversed. In other words, one traverses a boundary,
which means flipping the inside/outside status of the concerned operators $B(t)$.
$C_X$ denotes the complement of a set $X$. In terms of bit manipulation,
it simply means XORing the bitvectors $I(d_1)$ and $B(t)$.

If the map is composed of several connected components, then one needs
to do two different things for each connected component:
\begin{itemize}

\item \emph{Find a dart $d$ on the external boundary of the connected
component.} To do so, among all the shells $<\sigma_1,\alpha_2>$ in
  the connected component, we select the one that encloses the largest
  volume. It is found in linear time;
\item \emph{compute $I(d)$.} The connected component may be an internal
  boundary (for instance if you compute the difference between two
  concenric balls). It could be also included inside another object.
  It is not possible to deduce $I(d)$ from the sole combinatorial information,
  it requires some geometric tests: first, $I(d)$ is initialized to $\emptyset$. Then,
  a ray is launched from a point in $t(d)$,  and the inside/outside status of the operators $B(t)$ arre flipped
  for each intersected triangle $t$ (again, in terms of bit manipulation, this simply means
  XOR-ing $I(d)$ with $B(t)$. \new{In terms of predicates, ray-triangle intersection uses {\tt orient\_3d},
    computed using the coordinates of the input points and the exact coordinates of the intersections,
    (more on this in \secref{sec:arithmetics}), therefore the ``ray leakage'' phenomenon observed when using
    approximate coordinates cannot occur. However, the configuration where a ray traverses exactly an edge of the triangle
    requires special handling. I tested two approaches. The first one uses symbolic perturbation \citep{DBLP:journals/corr/EdelsbrunnerM94a}. The second one (less elegant but much simpler) keeps launching rays with random directions until no
    degeneracy is encountered. Similar performance was obtained (remember that only one ray launching per connected component
    of the volumetric mesh is required). I recommend the second approach (though less elegant, it is far simpler to implement).
  }
\end{itemize}

To summarize, using the Weiler model, classification is mostly a combinatorial
operation. Geometric computation is needed only in two places:
\begin{itemize}
  \item radial-sorting one bundle per radial polyline;
  \item tracing one ray per connected component
\end{itemize}

  \begin{figure}
    \centerline{\includegraphics[width=0.8\columnwidth]{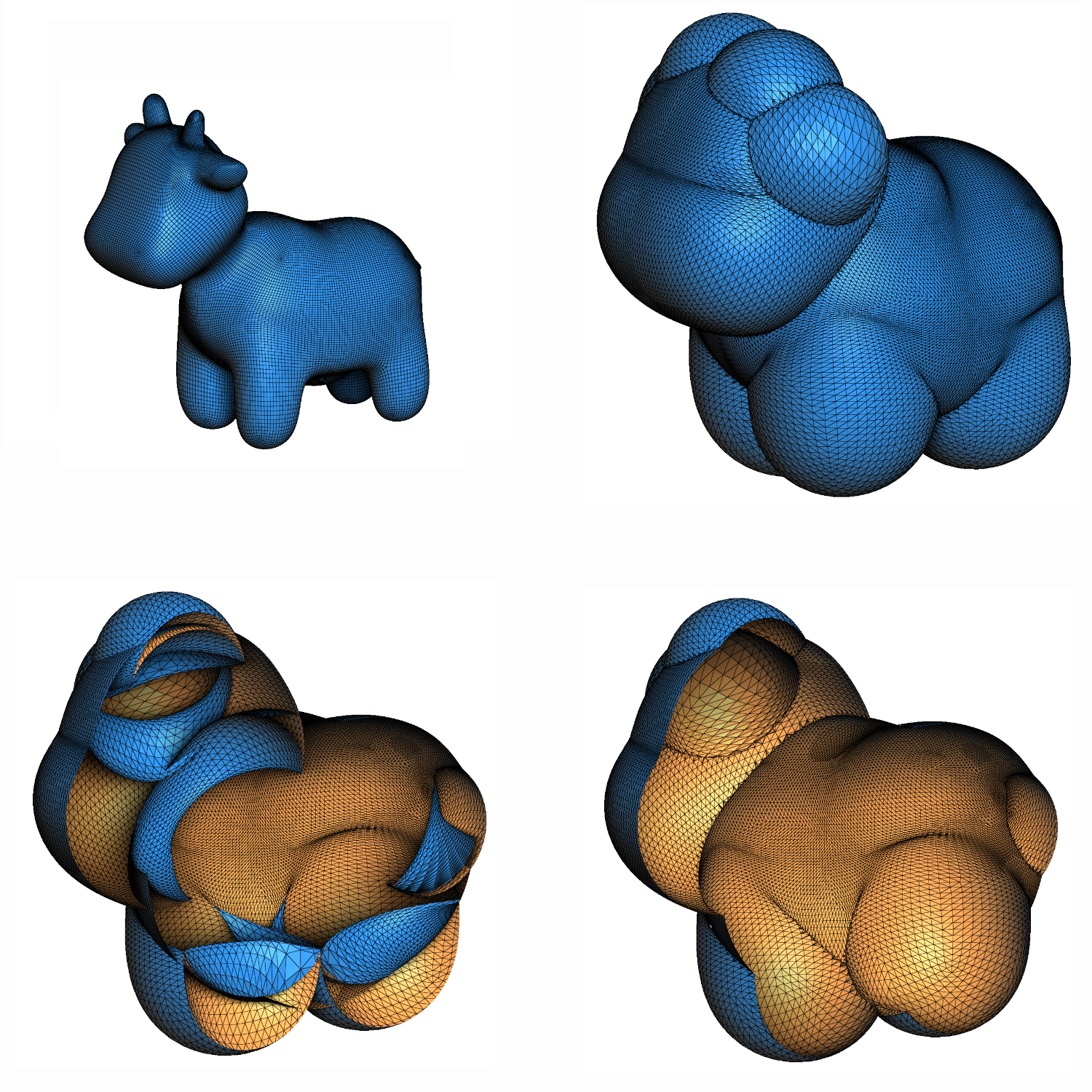}}
    \caption{Using the Weiler model to remove the internal garbage after a (naive) offsetting operation.}
    \label{fig:offset}
  \end{figure}

\paragraph*{\bf Other mesh repair operation: remove internal ``garbage''}

Besides providing an efficient combinatorial data structure for
implementing arbitrary boolean operations, the Weiler model can also
be used to implement other mesh repair operations. For instance, in 3D
mesh generation, one often wants to extract the ``outer skin'' from a
(possibly self-intersecting) polygon soup. As shown in Figure \ref{fig:offset},
one may also need to remove the internal ``garbage'' after performing
a naive offsetting operation (here all vertices were simply moved a certain
distance along their normal vectors). To do so, one possibility is to
compute the union of everything, however, if the input is a polygon
soup, it will not be easy to tag each individual primitives. Another
possibility is to extract the outer shell of all connected component
and keeping the ones that are not included in other components.
This operation is trivial to implement from the two algorithms
of the previous paragraph (the one that determines the outer shell and
the one that traces a ray).

\subsubsection{Simplifying}
\label{sec:simplify}

  \begin{figure}
    \centerline{\includegraphics[width=\columnwidth]{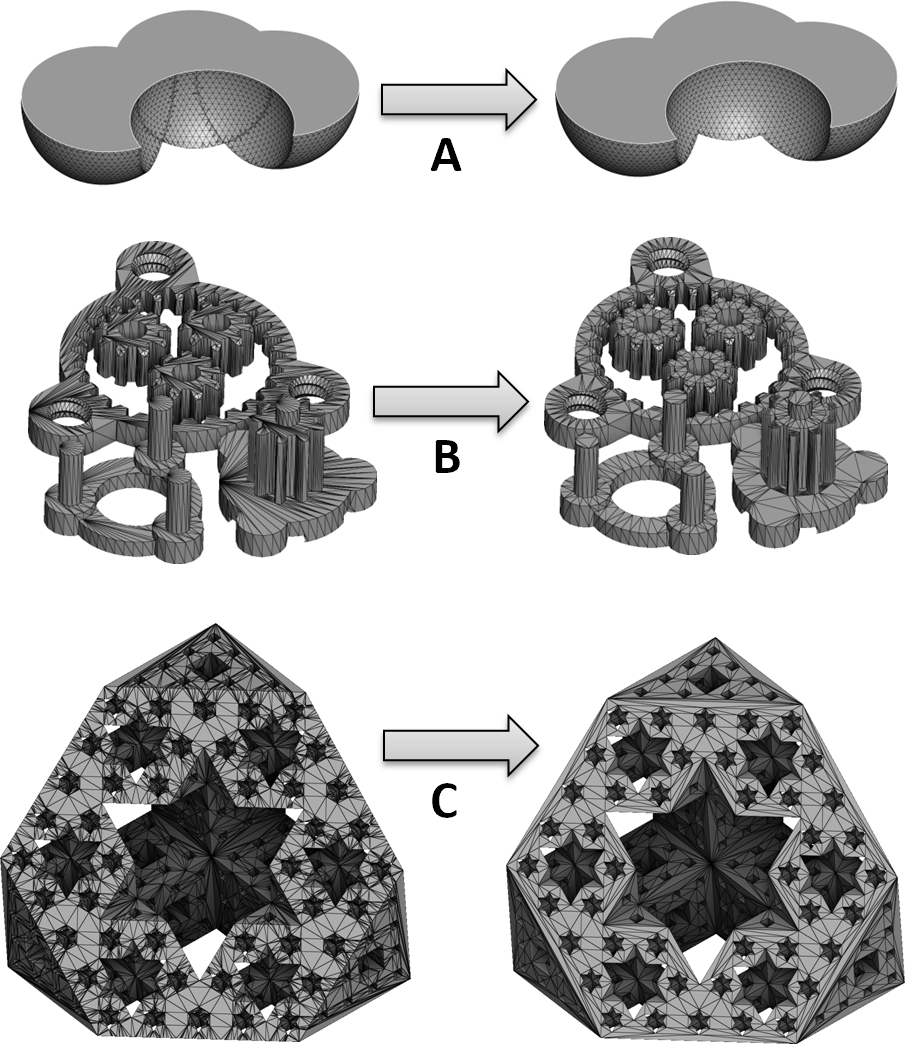}}
    \caption{Simplification of the coplanar facets}
    \label{fig:simplify}
  \end{figure}

The last step of our pipeline is the simplification of co-planar
facets, (step 5 in Figure \ref{fig:CSG_overview}
P. \pageref{fig:CSG_overview}), recalled in  Figure
\ref{fig:simplify}-A: the border of the volume defined by the boolean
expression may present the ``scars'' of intersection that were
computed during the co-refinement phase, that are not necessarily
needed in the final mesh.  One possibility to remove these scars would
be to keep in each triangle of the co-refinement a reference to the
input triangle, so that triangles coming from the same initial
triangle could be merged in a post-processing phase. However, the
input data may also contain poorly triangulated planar areas, worth remeshing,
such as
the example shown in Figure \ref{fig:simplify}-B). The regions composed
of co-planar facets are detected by greedily traversing them, using an
exact predicate that tests the co-linearity between the normal vectors of two adjacent facets.
Then, the borders of these regions are extracted. Finally,
the edges of the borders are inserrted in a constrained
Delaunay triangulation, using the same algorithm (and the same code)
as in the co-refinement phase. Finally, the triangles in the external
zone and in the (potential) internal pocket are discarded thanks to a
greedy traversal.

In addition, before inserting the border edges into the constrained
Delaunay triangulation, the borders can be
simplified: one can discard a vertex of the border provided that it is
aligned with its predecessor and successor along the border and
provided that it does not appear somewhere else in the mesh. The effect of
this simplification is shown in Figure \ref{fig:simplify}-C.

\begin{figure*}
\centerline{\includegraphics[width=0.95\textwidth]{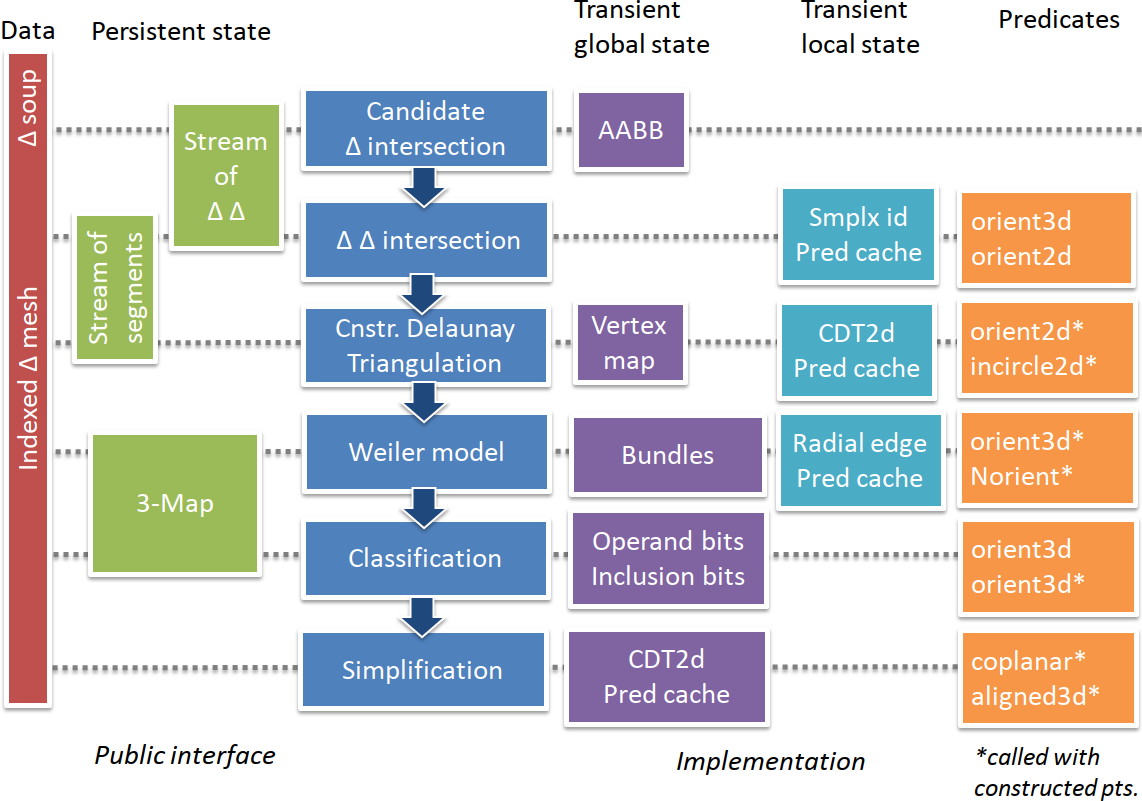}}
\caption{Global architecture of the algorithm}
\label{fig:architecture}
\end{figure*}

\subsection{Summary of the algorithmic pipeline architecture}

Before diving into the details of the arithmetic kernels, let us take a step backwards
and see how the different components mesh together. As mentioned in the introduction,
there are many possible choices for the data structure that store and share combinatorial
information between the stages of the pipeline, and one needs to find the right balance:
on the one hand, storing more combinatorial information may improve performance. On the
other hand, introducing more data structure makes the system more complicated, and more
difficult to maintain. The choices that I made are summarized in Figure \ref{fig:architecture}.
The data shared by all stages of the pipeline is simply an indexed triangulated mesh. In addition,
some stages share a \emph{persistent stage}. For instance, the AABB sends a stream of candidate
triangle intersection pairs to the triangle-triangle intersection, that in turn sends a stream
of segments to the constrained Delaunay triangulation. The Weiler model and the classification phase
share a 3-map. This defines the public interface of the pipeline stages (left part of the figure).
Each stage has an internal state (right part of the figure), that disappears at the end of the
stage (transient). A part of it is global in the mesh, and a part of it is more local,
attached to individual elements (or small group of elements) of the mesh. Finally, each stage
uses some predicates, some of them applied to the original vertices, and some of them to the
computed intersections. In this architecture, the combinatorial information communicated between
the stages is very reduced, which makes it easier to develop, unit-test and maintain each indivual
stage. It is made possible by pushing a significant part of the complexity towards the exact
representation of the points and the predicates, detailed in the next section.

\subsection{Nuts and bolts: two arithmetic kernels}
\label{sec:arithmetics}

I shall now give more details about the arithmetic kernels.
The reader may skip this section in a first read.
This section contains details that I considered worth sharing, about how to implement
arbitrary precision numbers and about how to derive the formulas for the predicates
acting on them, together with non-trivial details that have important consequence on
performance.
I tested two different
versions, one based on arithmetic expansions, and the other one based on multiprecision
floating point arithmetics. Each kernel provides an \verb|exact_nt| number type, that
supports addition, subtraction and product, 2d and
3d vectors with cartesian and homogeneous coordinates represented by \verb|exact_nt|, as
well as the standard predicates \verb|orient_2d|, \verb|orient_3d|,
and symbolically perturbed \verb|in_circle|. Each kernel also provides lexicographic
points comparison, used to create the global vertex table (see Section \ref{sec:CDT2d}) and
used by the symbolic perturbations.
For both kernels, I detail how numbers are represented, and
how the different predicates are implemented.

\subsubsection{The expansion-based arithmetic kernel}
\label{sec:expansions}

\paragraph*{\bf Arithmetic expansions}

With arithmetic expansions (explained in great detail in
\cite{shewchuk96a,shewchuk97a} and
mentioned in Section \ref{sec:previous_work}), each number is represented by an
array of floating-point numbers called \emph{components}.  The
represented number corresponds to the sum of the components.
In addition, these components are
sorted by decreasing exponents, and satisfy a special property: they
are \emph{non-overlapping}, that is, the exponents and the bits sets
in the mantissa are such that the sign is determined by the first
component. Arithmetic expansions are based on the \emph{two-sum}
algorithm \cite{10.1007/BF01975722}, or its fast version \cite{knuth97},
that computes the rounded sum $x_1$ of two numbers $a$ and $b$,
and the exact round-off error $x_2$, as follows:
$$
\begin{array}{lcl}
  x_1 & \leftarrow & a \oplus b \\
  x_2 & \leftarrow & b \ominus (x_1 \ominus a).
\end{array}
$$
where $\oplus$ and $\ominus$ denote the addition and subtraction of
IEEE754 floating point numbers rounded to nearest.

By properly orchestrating these operations (as well as a more complicated
\emph{two-prod} operation that computes the same information for products),
one can implement addition, subtraction and product for expansions of
arbitrary lengths, and use them to implement exact
geometric predicates, as explained by Shewchuk
in \cite{shewchuk96a,shewchuk97a}. The \emph{distillation} operation required
to compute the product of expansions of arbitrary lengths, also evoked in
the references above, is implemented in \cite{DBLP:journals/cad/Levy16}.

The idea here is to test whether arithmetic expansions can be used to
implement \emph{exact constructions}. In our case, these exact
constructions are used to compute the new intersection points. Note
also that some predicates take these constructed points as inputs, in
the constrained Delaunay triangulations and in the radial sort. This means
that the algebraic operations on the constructed point's coordinates are going
to be chained. This chaining / nesting of expressions that is much deeper
than with the classical usage of arithmetic expansions has two consequences:

\begin{itemize}

\item with arithmetic expansions, the representation of the same
  number is non-unique: in the extreme case, one could imagine using a
  single component for each non-zero bit in the number. If not enough care
  is taken, the length of the stored numbers grow larger and larger, as well
  as computation time;

\item arithmetic expansions are not limited in length, but it is not sufficient
  to ensure that arbitrary numbers can be manipulated: exponents are stored
  with a limited number of bits (11 bits in double precision), and overflow and
  underflow may occur when multiplying very large or very small numbers.
  This limit is quickly reached, for instance, when using the \verb|in_circle|
  predicate on points coming from the intersection of several
  triangles.

\end{itemize}

Both bottlenecks can be mitigated as follows:

\paragraph*{\bf Compression}

Compression is an operation that takes an expansion and that
returns a more compact expansion with the same value.
It is described in Section 2.8 of \cite{shewchuk97a}, and can
be summarized as in the algorithm below\footnote{
I think there is a typo in the original article,
line 14 should read $h_{top} \leftarrow q$ (small $q$ and not capital $Q$).
I think that this error was overlooked before because compression is mostly needed
when cascading operations, as done here since expansions are used in exact constructions,
and it was probably not done before.
}.

$$
\begin{array}{ll}
  &\mbox{compress(e)} \\
  &\mbox{\bf input/output:}\quad  e:\mbox{an expansion (compressed in-place)} \\
  \hline
  \\
  (0)  & m \leftarrow \mbox{length}(e) \\
  (1)  & Q \leftarrow e_m \\
  (2)  & bottom \leftarrow m \\
  (3)  & \algfor i = m-1 \algto 1 \\
  (4)  & \quad (Q, q) \leftarrow \mbox{fast\_two\_sum}(Q, e_i) \\
  (5)  & \quad \algif q \neq 0 \algthen \\
  (6)  & \quad \quad e_{bottom} \leftarrow Q \\
  (7)  & \quad \quad bottom \leftarrow bottom - 1 \\
  (8)  & \quad \algend \\
  (9)  & \algend \\
  (10) & e_{bottom} \leftarrow Q \\
  (11) & top \leftarrow 1 \\
  (12) & \algfor i = bottom+1 \algto m \\
  (12) & \quad (Q, q) \leftarrow \mbox{fast\_two\_sum}(e_i,Q) \\
  (13) & \quad \algif q \neq 0 \algthen \\
  (14) & \quad \quad e_{top} \leftarrow q \\
  (15) & \quad \quad top \leftarrow top + 1 \\
  (16) & \quad \algend \\
  (17) & \algend \\
  (18) & e_{top} \leftarrow Q \\
  (19) & \mbox{set\_length}(e, top)
\end{array}
$$

Compression proceeds by sweeping the expansion twice, in both
directions, first from largest to smallest component, then from
smallest to largest, ``swallowing'' a component each time the rounded sum
of two successive components is exact (test $q\neq0$ line 7 and
16). Since the speed of the arithmetic operations dramatically depend
on the length of the involved expansions, this function is called
before each complicated operation, such as computing a determinant,
and before storing a constructed point. It is worth it, because expansion product
costs a lot (in $O(m \times n)$, product of expansion lengths).

\paragraph*{\bf Orientation predicates}

I shall now explain how to compute the different predicates that we need.
The orientation predicates (\verb|orient_2d| and \verb|orient_3d|) are
classical. The only subtlety is that the points are in homogeneous coordinates.
Given three points $\bp_0, \bp_1, \bp_2$ with homogeneous coordinates
($\bp_i = [x_i\ y_i\ w_i]$), the predicate \verb|orient_2d| writes:
$$
{\tt orient\_2d}(\bp_0, \bp_1, \bp_2) =
  \mbox{sign}(U_w) \mbox{ sign}(V_w) \mbox{ sign}
  \left|
  \begin{array}{cc}
  U_x & U_y \\
  V_x & V_y
  \end{array}
  \right|.
$$
where:
$$
\begin{array}{lcll}
  U & = & [x_1 - x_0\quad y_1 - y_0\quad w_1] & \mbox{if } w_1 = w_0 \\
  U & = & [w_0 x_1 - w_1 x_0\quad w_0 y_1 - w_1 y_0\quad w_0 w_1 ] & \mbox{otherwise}\\[2mm]
  V & = & [x_2 - x_0\quad y_2 - y_0\quad w_2] & \mbox{if } w_2 = w_0 \\
  V & = & [w_0 x_2 - w_2 x_0\quad w_0 y_2 - w_2 y_0\quad w_0 w_1 ] & \mbox{otherwise} \\
\end{array},
$$

and $\verb|orient_3d|$ is written similarly.

\emph{Note:} one could have used instead the alternative expression:
$$
{\tt orient\_2d}(\bp_0, \bp_1, \bp_2) =
\left|
  \begin{array}{ccc}
  x_0 & y_0 & w_0 \\
  x_1 & y_1 & w_1 \\
  x_2 & y_2 & w_2
  \end{array}
\right|
$$
but it is in general preferable to use expressions with coordinate differences,
leading to much smaller expansions (see \cite{shewchuk96a,shewchuk97a}). The effect
is even more dramatic in our case, where input points come from exact constructons.

All computations are done using arithmetic expansions.
To speed-up computation in the easy cases, a
filter based on interval arithmetics is used. To convert an expansion $e$
into an interval, that is, finding the tightest interval that contains the
exact number represented by $e$, I add all components of $e$ to the interval
in decreasing magnitude order, stop as soon as next component is smaller than
ulp, then expand interval by 1 ulp.

  \begin{figure}
    \centerline{\includegraphics[width=\columnwidth]{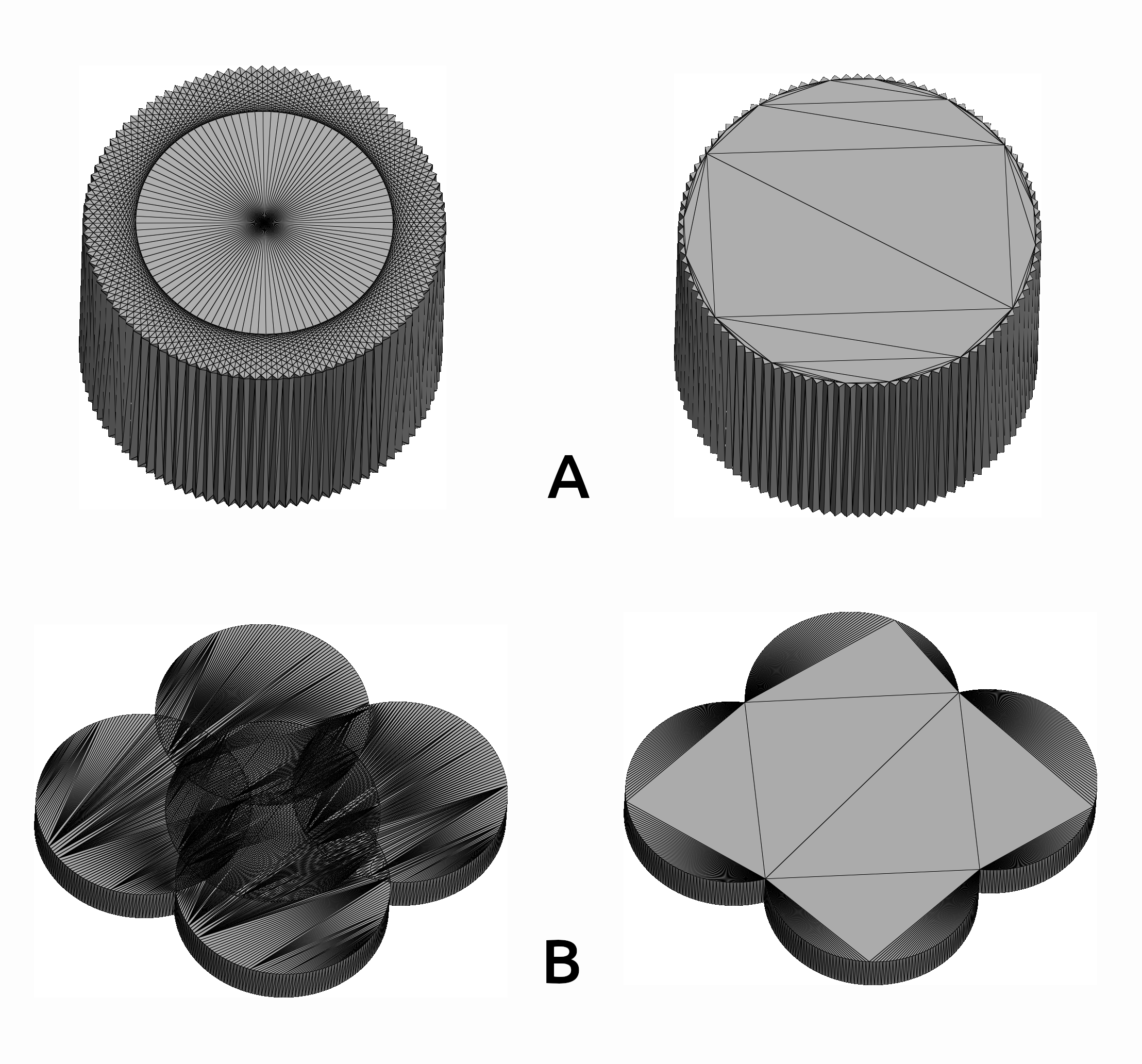}}
    \caption{Examples of mesh unions computed with arithmetic expansions. A: union of 50 rotated cubes. B: union of four cylinders, with many coplanar facets. Both examples are computed without (left) and with (right) simplification of coplanar facets.}
    \label{fig:expansions}
  \end{figure}

\paragraph*{\bf The in-circle predicate and its symbolic perturbation}
The in-circle predicate is more subtle, because it has higher degree.
Clearly one could directly implement the classical formula (written here
with cartesian coordinates):
$$
{\tt in\_circle}(\bp_0, \bp_1, \bp_2, \bp_3) = \mbox{sign}\left|
\begin{array}{cccc}
  x_0 & y_0 & x_0^2 + y_0^2 & 1 \\[1mm]
  x_1 & y_1 & x_1^2 + y_1^2 & 1 \\[1mm]
  x_2 & y_2 & x_2^2 + y_2^2 & 1 \\[1mm]
  x_3 & y_3 & x_3^2 + y_3^2 & 1
\end{array}
\right|.
$$
However, we remind that the higher degree in this expression can be a problem in our case,
because our input points
come from constructions, and nesting too many operations with arithmetic
expansions can lead to overflow or underflow due to the limited number of
bits to represent an exponent. The multi-precision kernel presented
in the next section does not have this limitation, but let us see whether this
issue can be mitigated with arithmetic expansions.
There exists a more general version of the \verb|in_circle| predicate, used to construct
power diagrams (also called Laguerre diagrams) and their duals
(called regular triangulations). This predicate takes additional weights
$\psi_0, \psi_1, \psi_2, \psi_3$ as arguments and writes:
$$
\begin{array}{l}
{\tt in\_circle\_weighted}(\bp_0, \bp_1, \bp_2, \bp_3, \psi_0, \psi_1, \psi_2, \psi_3)
= \\[2mm] \mbox{sign}\left|
\begin{array}{cccc}
  x_0 & y_0 & x_0^2 + y_0^2 - \psi_0 & 1 \\[1mm]
  x_1 & y_1 & x_1^2 + y_1^2 - \psi_1 & 1 \\[1mm]
  x_2 & y_2 & x_2^2 + y_2^2 - \psi_2 & 1 \\[1mm]
  x_3 & y_3 & x_3^2 + y_3^2 - \psi_3 & 1
\end{array}
\right|.
\end{array}
$$

So imagine now that you rewrite the \verb|in_circle| predicate and pass
to it additional arguments $l_i$ as follows:
$$
{\tt in\_circle\_l}(\bp_0, \bp_1, \bp_2, \bp_3, l_0, l_1, l_2, l_3) = \mbox{sign}\left|
\begin{array}{cccc}
  x_0 & y_0 & l_0 & 1 \\[1mm]
  x_1 & y_1 & l_1 & 1 \\[1mm]
  x_2 & y_2 & l_2 & 1 \\[1mm]
  x_3 & y_3 & l_3 & 1
\end{array}
\right|.
$$
If you use $l_i = x_i^2 + y_i^2$ \new{(computed exactly)}, then you obtain exactly the same result as $\tt in\_circle$.
\new{Now, consider that you replace the (exact) value of $l_i$ with the nearest floating-point number, that is,}
$l_i = \mbox{round\_to\_neareset}(x_i^2 + y_i^2)$, then what you
obtain is the same result as $\tt in\_circle\_weighted$, with
$\psi_i = \mbox{round\_to\_neareset}(x_i^2 + y_i^2) - (x_i^2 + y_i^2)$. While it is not exactly
the same result as $\tt in\_circle$, what you obtain in the end is still a well-defined object,
a Regular Triangulation (dual of power diagram), very similar to the Delaunay triangulation,
except for a few flipped edges, and more importantly, provided that always the same $l$ is used
for the same point, and with adapted symbolic perturnation, the triangulation remains uniquely defined.
There is however an important difference: if \verb|in_circle_l|$(\bp_1,\bp_2,\bp_3,\bp_4,l_1,l_2,l_3,l_4)$
is negative, it does not necessarily imply that $(\bp_1,\bp_2,\bp_3,\bp_4)$ forms a convex quadrilateral.
In the constrained Delaunay triangulation, this condition needs to be explicitly tested before flipping an edge.

We shall now see how to write the predicate for points
with homogeneous coordinates and the associated symbolic perturbation.

Take the determinant in \verb|in_circle_l| above, subtract the last row from the first three rows
then develop w.r.t. the fourth row:
$$
\begin{array}{lcl}
  {\tt in\_circle\_l}
    & = & \mbox{sign}\left|
\begin{array}{cccc}
  x_0 - x_3 & y_0 - y_3 & l_0 - l_3 \\[1mm]
  x_1 - x_3 & y_1 - y_3 & l_1 - l_3 \\[1mm]
  x_2 - x_3 & y_2 - y_3 & l_2 - l_3 \\[1mm]
\end{array}
\right|.
\end{array}
$$

We now need to take into account that the points are represented with homogeneous coordinates, and
we need to find an expression without any division.

$$
\begin{array}{l}
  \mbox{using}:
\left\{
\begin{array}{lcl}
  (X_{i+1}, Y_{i+1}, W_{i+1}) & = & \bp_i - \bp_3 \quad \mbox{in homo. coords.} \\
  L_{i+1} & = & l_i - l_3
\end{array}
\right.\\ \\
\mbox{one gets:} \quad {\tt in\_circle\_l} = \mbox{sign}\left|
\begin{array}{ccc}
  X_1 / W_1 & Y_1 / W_1 & L_1 \\
  X_2 / W_2 & Y_2 / W_2 & L_2 \\
  X_3 / W_3 & Y_3 / W_3 & L_3
\end{array}
\right|.
\end{array}
$$

Developping along the last column, factoring the $W_i$'s out, and multiplying everything by $W_1 W_2 W_3$,
one finally obtains:
$$
\begin{array}{l}
{\tt in\_circle\_l} = \mbox{sign}(W_1) \times \mbox{sign}(W_2) \times \mbox{sign}(W_3) \times \\[3mm]
\quad \quad \mbox{sign} \left(
{L_1 W_1} \left| \begin{array}{cc} X_2 Y_2 \\ X_3 Y_3 \end{array} \right| \ -
{L_2 W_2} \left| \begin{array}{cc} X_1 Y_1 \\ X_3 Y_3 \end{array} \right| \ +
{L_3 W_3} \left| \begin{array}{cc} X_1 Y_1 \\ X_2 Y_2 \end{array} \right| \right).
\end{array}
$$

I am using Simulation of Simplicity \cite{DBLP:journals/corr/EdelsbrunnerM94a} to consistently take a decision
when the quantity above is zero. I consider that the points $\bp_0, \bp_1, \bp_2, \bp_3$ are geometrically sorted,
which defines local indices $i_0, i_1, i_2, i_3$ (a permutation of $\{0,1,2,3\}$).
I am using for that the lexicographic order on the point's cartesian coordinates $x_i/w_i, y_i/w_i$
(more on this in the next paragraph). Now I consider that the lengths parameters $l_0, l_1, l_2, l_3$
are replaced with $l_0 + \epsilon^{i_0}, l_1 + \epsilon^{i_1}, l_2 + \epsilon^{i_2}, l_3 + \epsilon^{i_3}$ for a small
$\epsilon$. The perturbed predicate is then defined by the sign of the first non-zero coefficient of
$\epsilon^{i_k}$. They are easy to find, using the following expression of \verb|in_circle_l| and developping
it with respect to the third row and keeping only the coefficients in $\epsilon^{i_k}$:
$$
\begin{array}{lcl}
{\tt in\_circle\_l} & = &
\left|
\begin{array}{cccc}
  x_0 & y_0 & l_0 + \epsilon^{i_0} & 1\\
  x_1 & y_1 & l_1 + \epsilon^{i_1} & 1\\
  x_2 & y_2 & l_2 + \epsilon^{i_2} & 1\\
  x_3 & y_3 & l_3 + \epsilon^{i_3} & 1
\end{array}
\right|\\ \\
\quad \quad \quad = \quad \ldots \quad & + &
\epsilon^{i_0} \left| \begin{array}{ccc} x_1 & y_1 & 1 \\ x_2 & y_2 & 1 \\ x_3 & y_3 & 1 \end{array} \right| -
\epsilon^{i_1} \left| \begin{array}{ccc} x_0 & y_0 & 1 \\ x_2 & y_2 & 1 \\ x_3 & y_3 & 1 \end{array} \right| + \\ \\
&  &
\epsilon^{i_2} \left| \begin{array}{ccc} x_0 & y_0 & 1 \\ x_1 & y_1 & 1 \\ x_3 & y_3 & 1 \end{array} \right| -
\epsilon^{i_3} \left| \begin{array}{ccc} x_0 & y_0 & 1 \\ x_1 & y_1 & 1 \\ x_2 & y_2 & 1 \end{array} \right|
\end{array}.
$$

In homogeneous coordinates, the signs of the coefficients of $\epsilon^{i_k}$ can be computed as follows:
$$
\begin{array}{l}
\mbox{sign}\left|
\begin{array}{ccc}
  x_1/w_1 & y_1/w_1 & 1 \\
  x_2/w_2 & y_2/w_2 & 1 \\
  x_3/w_3 & y_3/w_3 & 1
\end{array}
\right| = \\ \\
\mbox{sign}\left(
   w_1 \left| \begin{array}{cc} x_2 & y_2 \\ x_3 & y_3 \end{array} \right| -
   w_2 \left| \begin{array}{cc} x_1 & y_1 \\ x_3 & y_3 \end{array} \right| +
   w_3 \left| \begin{array}{cc} x_1 & y_1 \\ x_2 & y_2 \end{array} \right|
   \right) \ \times \\
            \mbox{sign}(w_1)
     \times \mbox{sign}(w_2)
     \times \mbox{sign}(w_3)
\end{array}
$$

\paragraph*{\bf Geometric sorting and geometric indexing}

The global vertex table (see Section \ref{sect:global_vertex_table}) and the symbolic perturbation introduced
above need a total order on the points. I am simply using the lexicographic order, based on
a function that compares the cartesian coordinates of two points. Since our points are stored
with homogeneous coordinates, it means comparing rationals. What we need is a new predicate:
$$
\begin{array}{l}
  {\tt ratio\_compare}(x_1,w_1,x_2,w_2)  = \mbox{sign}\left(\frac{x_1}{w_1} - \frac{x_2}{w_2}\right) \\
  \quad \quad =  \mbox{sign}(w_1) \times \mbox{sign}(w_2) \times \mbox{sign}\left( w_2 x_1 - w_1 x_2 \right).
\end{array}
$$

Since computing $w_2 x_1 - w_1 x_2$ takes significant time with long expansions, the predicate is optimized
in three particular cases where the result is trivial and where this computation can be avoided:
\begin{itemize}
  \item if $x_1$ and $x_2$ are both zero;
  \item if the signs of $x_1/w_1$ and $x_2/w_2$ differ;
  \item if $w_1 = w_2$.
\end{itemize}

The expansion-based arithmetic kernel works reasonably well in practice, and can be used for co-refinement
operations. The example shown in Figure \ref{fig:expansions} with many coplanar surfaces
demonstrates how it successfully generates a unique triangulation in them. For both example, the Euler-Poincaré
characteristic is 2, as expected. Hence, the \verb|in_circle_l| predicate makes it possible to push the limits
of what can be computed with arithmetic expansions. Co-refinements can be computed nearly with all models of the Thingi10K database. \\

However, one still reaches the limit when attempting to
create the Weiler model: the involved predicates compute cross-products between vectors joining constructed
points, and then dot product between those. Remember that the points are themselves the result of intersections.
I instrumented the code to output a histogram of the lengths of the expansions, and they can be as long as
65000. This is not that surprising, knowing that each operation can potentially double the length of the expansions,
this corresponds to 16 nested levels. Thanks to compression, this seldom occurs
(no more than a few times in multi-million element meshes). Besides the time and space requirement for these
very long expansions, a more important problem is that they can yield overflows and underflows when computing
products with them. For instance, using the expansion-based kernel, it can sucessfully evaluate the CSG
trees in \verb|example0001.csg| to \verb|example0020.csg| in the OpenSCAD category of
the ThingiCSG testsuite (see \ref{sec:thingiCSG} below), but it fails with all CSG trees
between \verb|example0021.csg| to \verb|example0024.csg|.

\subsubsection{The multiprecision floating-point arithmetic kernel}
\label{sec:mpfloat}

For this reason, and because the algorithm is used in
production by Yoyodine Corp\footnote{to be replaced with the real
company name in the final non-anonymous version}, I implemented and
tested an alternative kernel, that does not have the limitations of
the expansion-based kernel mentioned above.

\paragraph*{\bf Multiprecision floating-point arithmetics}

The kernel is based on multi-precision integers, implemented in the
GNU Multiple Precision library (GMP), similarly to what is done in
CGAL \cite{cgal:eb-23a}. As in CGAL, I represent a floating point
number with a mantissa $m$ stored as a multi-precision signed integer
from GMP (\verb|mpz_t|) and a 32 bits exponent $e$. The represented
number is $m \times 2^e$.  To ensure the uniqueness of the
representation, $m$ is constrained to have no trailing zero (its least
significant bit is always 1), with the exception of 0, always
represented as $0^0$.  GMP provides all the necessary functions
(initialization from integer, adding, subtracting, product, left and
right shift, comparisons). As compared to CGAL, I optimized some
operations, such as equality (compare sign, then exponent, then
mantissa only if they were the same), comparisons (easy answer if
signs differ), and comparison with special values such as 0 and 1. It has
a non-negligible impact in our context, where many operations are nested.

To convert a multiprecision floating-point
number into an interval, one first initializes both bounds of the
interval to the approximation of the number as a standard
floating-point number. If it did not fit in the floating-point number,
then the interval is enlarged by 1 ulp towards $-\infty$ or $+\infty$
depending on the sign of the number.

\paragraph*{\bf The predicates}

The orientation predicates use the same formulas as in the expansion-based kernel.
For the \verb|in_circle| predicate, one could reuse the formulas of the expansion-based
kernel, and inject the exact computation of $l_i = x_i^2 + y_i^2$ into them, however,
it is better to make \emph{difference of coordinates} appear in the computed determinants,
since it reduces cancellation errors in general, and improves the performance of the
arithmetic filter based on intervals. Let's start from the original expression of the
predicate, recalled here:

$$
{\tt in\_circle}  =
\mbox{sign} \left|
\begin{array}{cccc}
  x_0 & y_0 & x_0^2 + y_0^2 & 1\\
  x_1 & y_1 & x_1^2 + y_1^2 & 1\\
  x_2 & y_2 & x_2^2 + y_2^2 & 1\\
  x_3 & y_3 & x_3^2 + y_3^2 & 1
\end{array}
\right|.
$$

Then, you translate $\bp_3$ to the origin and develop with respect to the last row:
$$
{\tt in\_circle}
 =  \mbox{sign}
\left|
\begin{array}{ccc}
  x_0 - x_3 & y_0 - y_3 & (x_0 - x_3)^2 + (y_0 - y_3)^2 \\
  x_1 - x_3 & y_1 - y_3 & (x_1 - x_3)^2 + (y_1 - y_3)^2 \\
  x_2 - x_3 & y_2 - y_3 & (x_2 - x_3)^2 + (y_2 - y_3)^2 \\
\end{array}
\right|.
$$

This determinant is very similar to the one obtained in the
previous subsection, with the difference that the coefficients
in the third column are $(x_i - x_3)^2 + (y_i - y_3)^2$
instead of $(x_i ^2 + y_i^2) - (x_3^2 + y_3^2)$. It may be
surprising that both expression are equivalent, but remember
that the first one was obtained by row manipulations, and
the second one by geometric reasoning, both types of
transform leaving the determinant invariant.

Rewriting the determinant in terms of the homogeneous coordinates
$(X_i, Y_i, W_i)$ of $\bp_i - \bp_3$ and $L_i = X_i^2 + Y_i^2$, one gets:
$$
{\tt in\_circle}  = \mbox{sign}\left(\
\frac{1}{W_0^2 W_1^2 W_2^2} \left|
  \begin{array}{ccc}
    W_0 X_0 & W_0 Y_0 & L_0 \\
    W_1 X_1 & W_1 Y_1 & L_1 \\
    W_2 X_2 & W_2 Y_2 & L_2
  \end{array}
\right|\ \right).
$$

The positive factor can be dropped (we are only interested in the sign). Developing
w.r.t. the last column, one finally gets:
$$
  \begin{array}{l}
 {\tt in\_circle} = \mbox{sign}\left(\quad
 L_0 W_1 W_2 \left| \begin{array}{cc} X1 & Y1 \\ X_2 & Y2 \end{array} \right|\right. \quad - \\ \\
  \left. L_1 W_0 W_2 \left| \begin{array}{cc} X0 & Y0 \\ X_2 & Y2 \end{array} \right| \quad + \quad
  L_2 W_0 W_1 \left| \begin{array}{cc} X0 & Y0 \\ X_1 & Y1 \end{array} \right|\quad  \right)
\end{array}
$$

\paragraph*{\bf Geometric sorting and geometric indexing}

As with the ex\-pan\-sion-based kernel, we need a total order on
the points. Clearly we could use exactly the same formula as
what we did for expansions, but we can exploit the uniqueness
of the representation (that we did
not have with expansions). The representation of 3D points stored
with homogeneous coordinates $(x,y,z,w)$ is normalized as follows:
\begin{itemize}
\item $x,y,z,w$ are divided by their mutual gcd;
\item $w$ is positive;
\item the exponent of $w$ is zero.
\end{itemize}

With this convention, the representation of a point is unique, and
one can simply use the lexicographic order on $x,y,z,w$, forgetting
their geometric nature. Clearly, it will give a different order
as compared to the lexicographic order on $x/w,y/w,z/w$ used before,
but it is not a problem since the only thing we need to have
for the symbolic perturbation and for the global vertex table to work
is a total order. I also noticed that we are not obliged to pre-shift
the numbers so that $w$'s exponent is zero, instead of that we pass the
shifts to the comparison function, and shift by their difference only.
It makes both the spatial indexing and predicates significantly more efficient
(else they keep shifting the same numbers left and right).

\section{Tests}
\label{sec:tests}

\subsection{Thingi10K}
\label{sec:thingi10K}

\begin{table}
  \begin{tabular}{lllll}
    \hline
       & Cherchi  & \ \ \ Ours & \ Ours   & \ \ Ours \\
    ID & \ et.al. & expansions & multi prec. & multi prec. \\
       &          & Delaunay   & Delaunay & no Del.  \\
    \hline
    {\tt 252784}   & 104 & 580 & 89  & 78  \\
    {\tt 1016333}  & 868 & X   & 115 & 112 \\
    {\tt 55928}    & 87  & 298 & 46  & 37  \\
    {\tt 12368052} & 120 & 541 & 152 & 114 \\
    {\tt 498461}   & 19  & 123 & 21  & 16  \\
    {\tt 338910}   & 8   & 103 & 13  & 10  \\
    {\tt 252785}   & 24  & 106 & 16  & 13  \\
    {\tt 498460}   & 12  & 92  & 12  & 11  \\
    {\tt 242236}   & 50  & 18  & 24  & 20  \\
    {\tt 242237}   & 49  & 11  & 22  & 21  \\
    \hline
  \end{tabular}
  \caption{
    Timings (in seconds) for the 10 models
    from Thingi10K with the largest number of intersections
  }
  \label{table:thingi10k}
\end{table}

\begin{figure}
  \centerline{\includegraphics[width=\columnwidth]{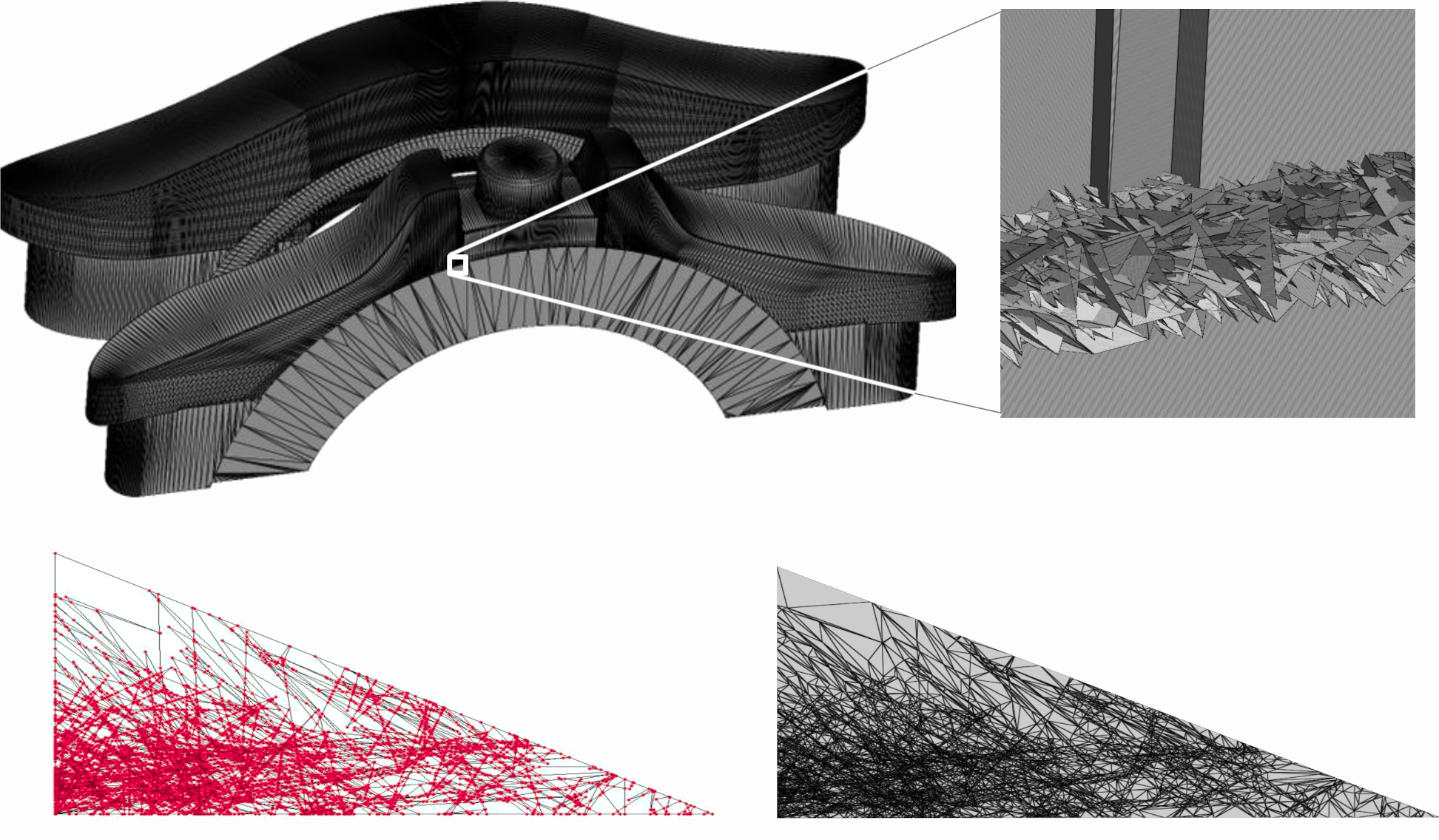}}
  \caption{One of Thingi10k's monsters, thing 996816.
    This mesh has a huge number of intersections,
    most of them located in the highlighted zone. It has up to several thousand
    intersections in the same triangle.}
  \label{fig:996816}
\end{figure}

\begin{figure}
  \centerline{\includegraphics[width=\columnwidth]{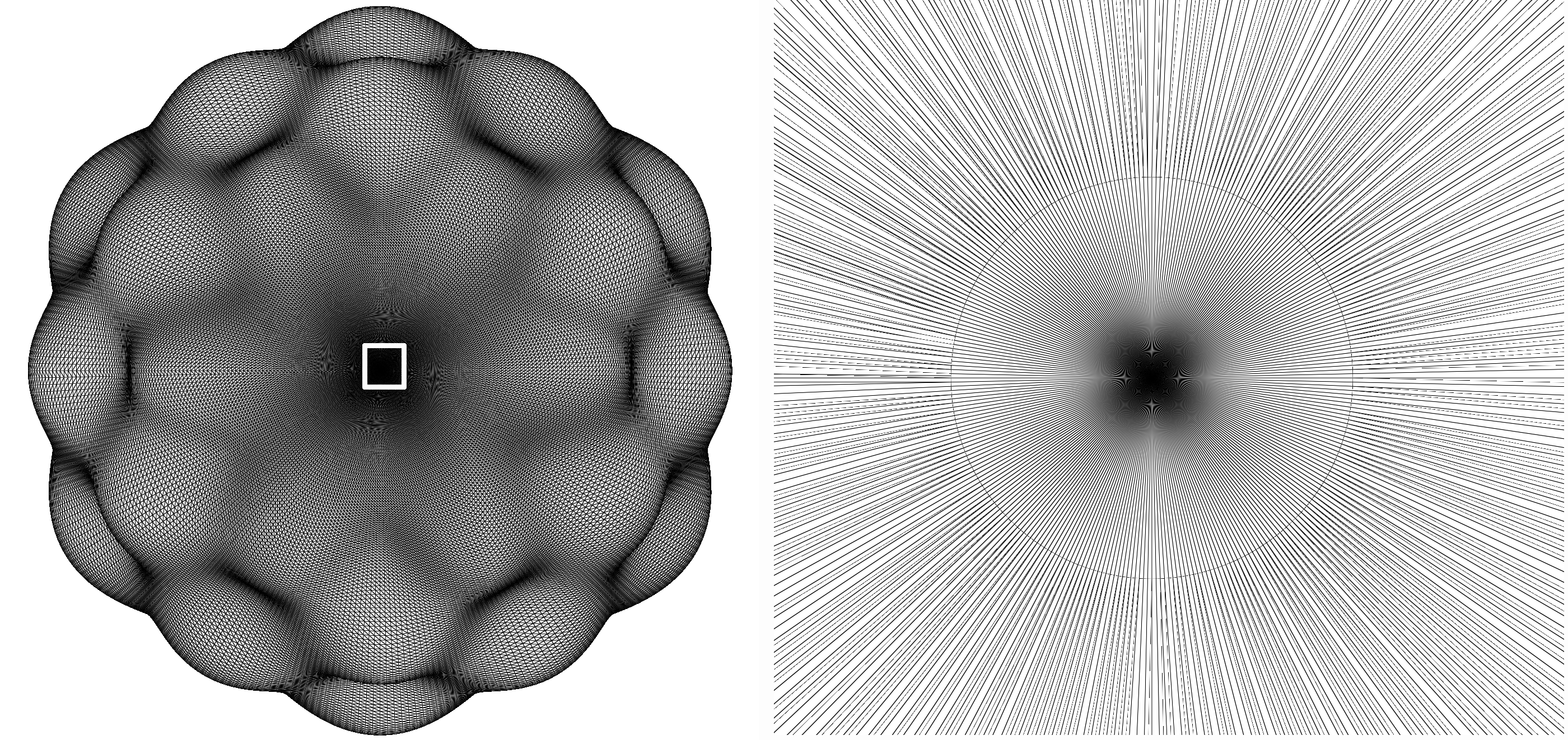}}
  \caption{Another monster from Thingi10k, thing 101633.
    This mesh stresses the arithmetic
    kernel a lot, with
    a large number of intersections located near a pole.}
  \label{fig:101633}
\end{figure}

I shall now report some timings and statistics, first with Thingi10K
\cite{Thingi10K}, a database with 10000 meshes. In the database, 4523
of them have intersections. The complete subset of models with
intersections is processed in 1h 45min. One of them (996816,
shown in Figure \ref{fig:996816}) is particular and takes 1271s (20 min)
to be processed. It has up to several thousands intersections in the
same triangle. This is because it has in a tiny zone a ``3D grid of triangles'',
that is, exactly the best method to create $O(N^3)$ intersections with $N$ triangles.
It is a good crash test for all the components of a mesh intersection algorithm,
in particular the constrained Delaunay triangulation.
Timings for the next 10 meshes with the largest number of intersections
are reported in Table \ref{table:thingi10k}. The
first column gives the timings for the state of the art
\cite{10.1145/3414685.3417818}. The second column corresponds to the
kernel based on arithmetic expansions. For one of the models (101633,
displayed in Figure \ref{fig:101633}), this kernel could not compute a
correct result, because underflows were encountered. It is explained
by the shape of the triangles that have intersections, that are very skinny
and intersecting near the pole. They generate arithmetic operations that
combine very large and very small numbers. This also explains the long
timing obtains with previous work on this model.
The third column
reports the timings obtained with the multi-precision floating point
kernel. As can be seen, for the largest models, timings are faster
than with \cite{10.1145/3414685.3417818}, and for some of them they
are slower. Faster timings are explained by our exact constructions:
in a certain sense, indirect predicates need to redo the same
computations several times, whereas exact constructions act as a
``cache''. Slower timings are explained by the constrained
\emph{Delaunay} triangulations that I compute, that involve the rather
costly \verb|in_circle| predicate. As far as 101633 is concerned, carefully
designing the arithmetic kernel and the associated predicates as done
in Section \ref{sec:mpfloat} has a significant impact on the performance.
The fourth column reports the
timings obtained with the multi-precision floating point kernel and
constrained triangulations (not Delaunay). It lets us see much
it costs to ensure the Delaunay property. With this kernel, timings
are almost always faster as compared to previous work.
Note that we loose the uniqueness of the
triangulation, and therefore one would need a pocket identification
algorithm as in \cite{10.1145/3414685.3417818} to make a fair
comparison. \new{To further compare both algorithms, that is, \cite{10.1145/3414685.3417818} and our method with constrained Delaunay triangulation, I conducted additional experiments, described in the next subsection.}

\begin{figure}
  \includegraphics[width=\columnwidth]{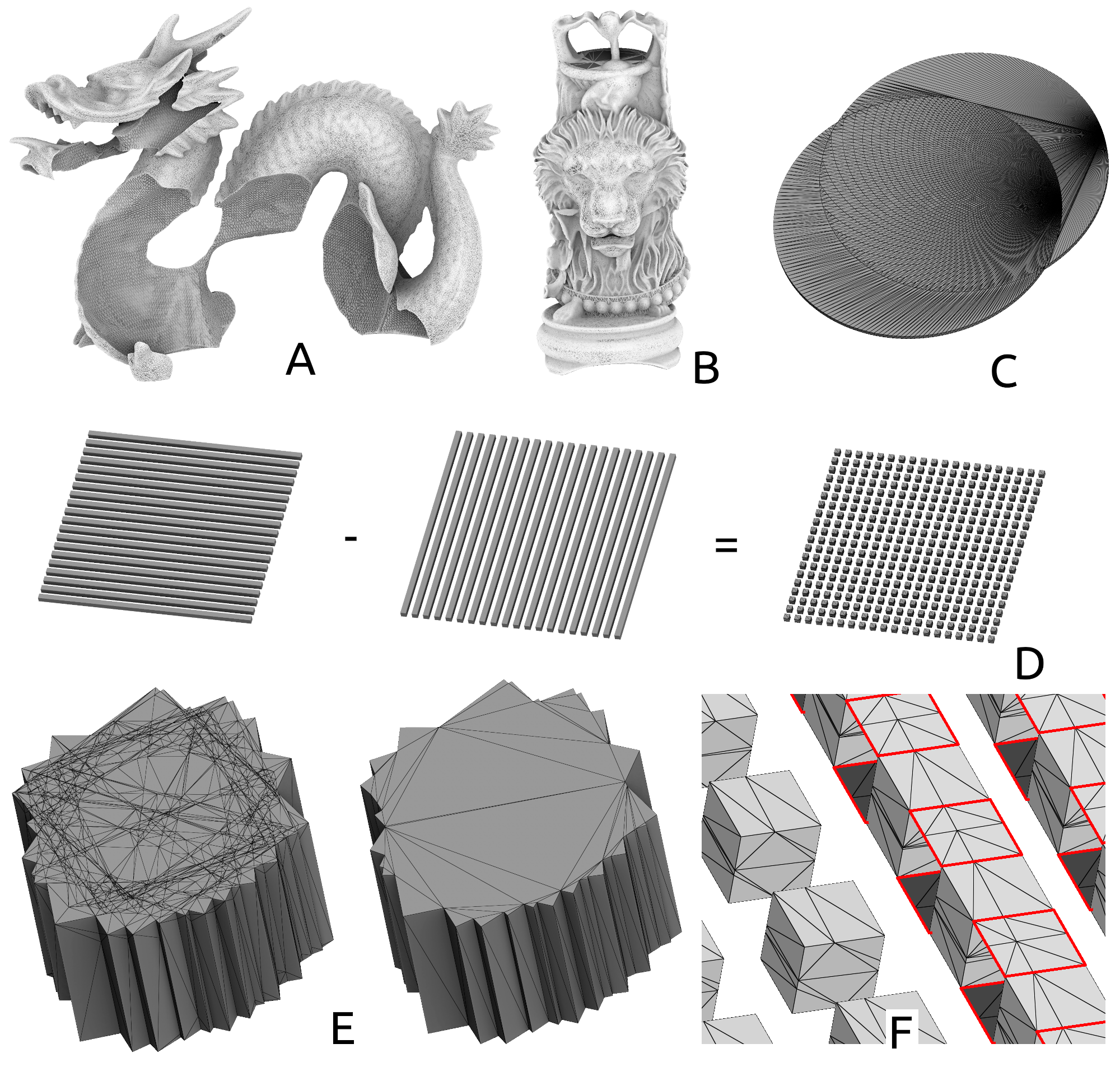}
  \caption{\new{Boolean operations with large triangle meshes and with highly degenerate configurations.
      The bottom one (E) is shown without and with simplification of co-planar facets. (F) shows the correct output of (D) on the left part, and the result obtained with Cherchi et.al's method on the right part, with spurious facets.
  }}
  \Description{Boolean operations with large triangle meshes and with highly degenerate configurations.}
  \label{fig:new_tests}
\end{figure}

\begin{table}
  \begin{tabular}{llll}
  \hline
  ID & ours & Zhou et.al & Cherchi et.al  \\
  \hline
  \verb|dragon-bunny|(A)  &  2.7 & 3.9  & 1.3 \\
  \verb|buddhaUlion|(B)   & 11.5 & 16.1 & 5.8 (sometimes crashes) \\
  \verb|cylUcyl|(C)       &  3.7 & 25.8 & 28.2 \\
  \verb|20rods-20rods|(D) &  0.9 & 3.3  & 2.1 (incorrect result) \\
  \hline
  \end{tabular}
  \caption{\new{
      Comparison with \cite{10.1145/2897824.2925901} and \cite{10.1145/3550454.3555460}, timings in seconds.
      Meshes are shown in Figure \ref{fig:new_tests}.
    }
  }
  \label{table:new_tests}
\end{table}


\begin{table*}
  \begin{tabular}{ll|llllllll|l|l}
  \hline
  ID & & AABB         & Box $\cap$ Box & $\Delta \cap \Delta$ & CDT & Radial & Weiler model  & Classification & Misc. & Total & Simplify \\
     & & construction &                &                      &     & sort   & combinatorics &                &       &       & coplanar \\
  \hline
  \verb|dragon-bunny| & (A)  & 106 & 172 & 1420 &  162 &  2 &  302 & 241 &  295 &  2700 & 10 \\
  \verb|buddhaUlion|  & (B)  & 512 & 811 & 6323 &  471 &  3 & 1969 &  10 & 1401 & 11500 & 24 \\
  \verb|20rods-20rods|& (C)  & <1  &   1 &   14 &  717 & 20 &   40 &  15 &   54 &   864 & 72 \\
  \verb|cylUcyl|      & (D)  &  1  &   6 &  183 & 3429 & <1 &   20 &   1 &   20 &  3660 & 57 \\
  \verb|rot_cube_20|  & (E)  & <1  &  <1 &   <1 &  646 & 23 &   36 &   1 &    3 &   712 & 15 \\
  \hline
  \end{tabular}
  \caption{\new{Timing breakdown (in milliseconds). Meshes are shown in Figure \ref{fig:new_tests}.
}
  \label{table:breakdown}
  }
\end{table*}

\subsection{Comparison with Zhou \emph{et.al} 2016 and Cherchi \emph{et.al} 2022}
\label{sec:CherchiZhou}

{\color{black}
In this section, I compare the new boolean operation algorithm with
\cite{10.1145/2897824.2925901} and \cite{10.1145/3550454.3555460}
that employ a very similar algorithm. Typical examples are reported in
Table \ref{table:new_tests} (the corresponding meshes are shown in Figure \ref{fig:new_tests}).
Examples A and B (scanned meshes with many small triangles, in generic position)
are taken from \cite{douze:hal-01121419}. Example C (union between two ``camembert cheese'' with
many co-planar small triangles) is inspired by \cite{10.1145/3550454.3555460} (Fig. 3). Examples
D and E are inspired by \cite{10.1145/3528223.3530181} (Figs 24 and 25).
For meshes with many tiny triangles in generic position, as in the \verb|dragon-bunny| (A)
and \verb|buddhaUlion| (B) test cases, the new method is faster than \cite{10.1145/2897824.2925901} and
slower than \cite{10.1145/3550454.3555460}. For meshes with many co-planar intersections,
such as in the \verb|cylUcyl| (C) and \verb|20rods-20rods| (D) test cases, the new method is up to $6x$ faster than
previous work. Cherchi et.al's method crashes unpredictably on the \verb|buddhaUlion| test case,
and produces for \verb|20rods-20rods| an incorrect result, with spurious hanging facets (shown in red in the right part of Fig. \ref{fig:new_tests}-F, the left part corresponds to the expected result). \\

In Table \ref{table:breakdown}, I report the timing breakdown of the algorithm
executed on the same datasets, where the different columns of the table correspond to:
\begin{itemize}
\item AABB construction: construction of the Axis-aligned bounding box tree \secref{sec:candidates};
\item Box $\cap$ Box: determinate candidate intersecting facets by traversing the AABB \secref{sec:candidates};
\item $\Delta \cap \Delta$: determine triangle pairs intersections \secref{sec:tritri};
\item CDT: Constrained Delaunay Triangulation \secref{sec:CDT2d};
\item Radial sort: sort triangles around non-manifold edges \secref{sec:Weiler};
\item Weiler combinatorics: combinatorial part of Weiler model construction \secref{sec:Weiler};
\item Classification: find the boundary of the result from the co-refinement \secref{sec:classify};
\item Misc: this regroups detecting co-located vertices and flat facets in the input, re-ordering mesh elements
  for better locality and faster multithreading, and detecting intersections located exactly on input vertices;
\item Simplify coplanar: optional simplification of co-planar facets \secref{sec:simplify}.
\end{itemize}

As can be seen, the highly-optimized triangulation algorithm
introduced here takes a minimal amount of time (more consequent for D that has many co-planar intersections, but still 6x faster than previous work). This is explained by both the carefully written predicates
and the Constrained Delaunay Triangulation method that exploits the combinatorial information to minimize
invoking the predicates and that minimizes dynamic memory allocation, making it especially efficient in
a multithreading context. The algorithm makes a maximum use of the combinatorial information, hence radial
sorting is done a limited number of times, as in \cite{10.1145/2897824.2925901}, and raytracing is only
required once per connected component of the 3-Map to determine inclusion, in contrast with \cite{10.1145/3550454.3555460}
that uses one ray-tracing query per surface patch. This comes at a significant price for creating the combinatorial structure,
which takes up to 10\% of the total computation time. Optionally it is possible to merge co-planar facets, as shown
in Figure \ref{fig:new_tests}-E (see also Figure \ref{fig:simplify} in \secref{sec:simplify} and Figure \ref{fig:nasty_gear_1}). It introduces a negligible overhead, thanks one again to the optimized CDT. Removing
unnecessary vertices in flat zones is crucial when evaluating deep CSG trees, as in the next subsection.
Without this post-processing, the number of triangles would quickly explode when chaining boolean operations.
The different contributions and specific choices that I made
gain important performance for boolean operations appearing in CAD-like CSG trees, that nearly systematically comprise
highly degenerate configurations with many co-planar intersecting triangles. However, even in this type
of configuration, the integer-based method EMBER is still spectacularly more efficient: according to the timings
reported in \cite{10.1145/3528223.3530181}.
On the \verb|20rods-20rods| (Fig. 24 in the EMBER article) it takes no more than 4.5 ms, and on a configuration similar to
the \verb|rot_cube_20| (Fig. 25 in the EMBER article) it takes 5.9 ms.
}

\begin{table}
  \begin{tabular}{lllll}
    \hline
    ID & ours & CGAL & CGAL   & manifold \\
       &      & NEF  & coref. & \\
    \hline
\verb|00_WarmUpExercise|              & <1  &   1 &  1 & <1 \\
\verb|01_Newbie2Guru15min|            &  2  &  15 &  1 &  1 \\
\verb|03_TeachingScript...| & 15 & 108 & 29 &  2 \\
\verb|13_hyperboloid|                 &  3  &  25   &  8 &  1 \\
\verb|14_LightSaber|                  &  1  &  10   &  3 & <1 \\
\verb|15_FamilyTreePendant|           &  1  &   8   &  1 & <1 \\
\verb|16_Ring5|                       &  1  &  25   &  7 & <1 \\
\verb|17_Tree|                        & 131 &  1721 & 11 & 2  \\
\verb|19_LEDlamp|                     &  6  &   151 & 96 & 1  \\
\verb|20_ElectricCircuit...|          &  2  &    35 & 20 & 1  \\
\verb|21_BasWheel|                    &  6  &    21 & 6  & <1 \\
    \hline
  \end{tabular}
  \caption{Timings (in seconds) for the {\tt Presentation} collection of ThingiCSG.}
  \label{table:thingicsg2}
\end{table}

\begin{figure*}
  \centerline{\includegraphics[width=\textwidth]{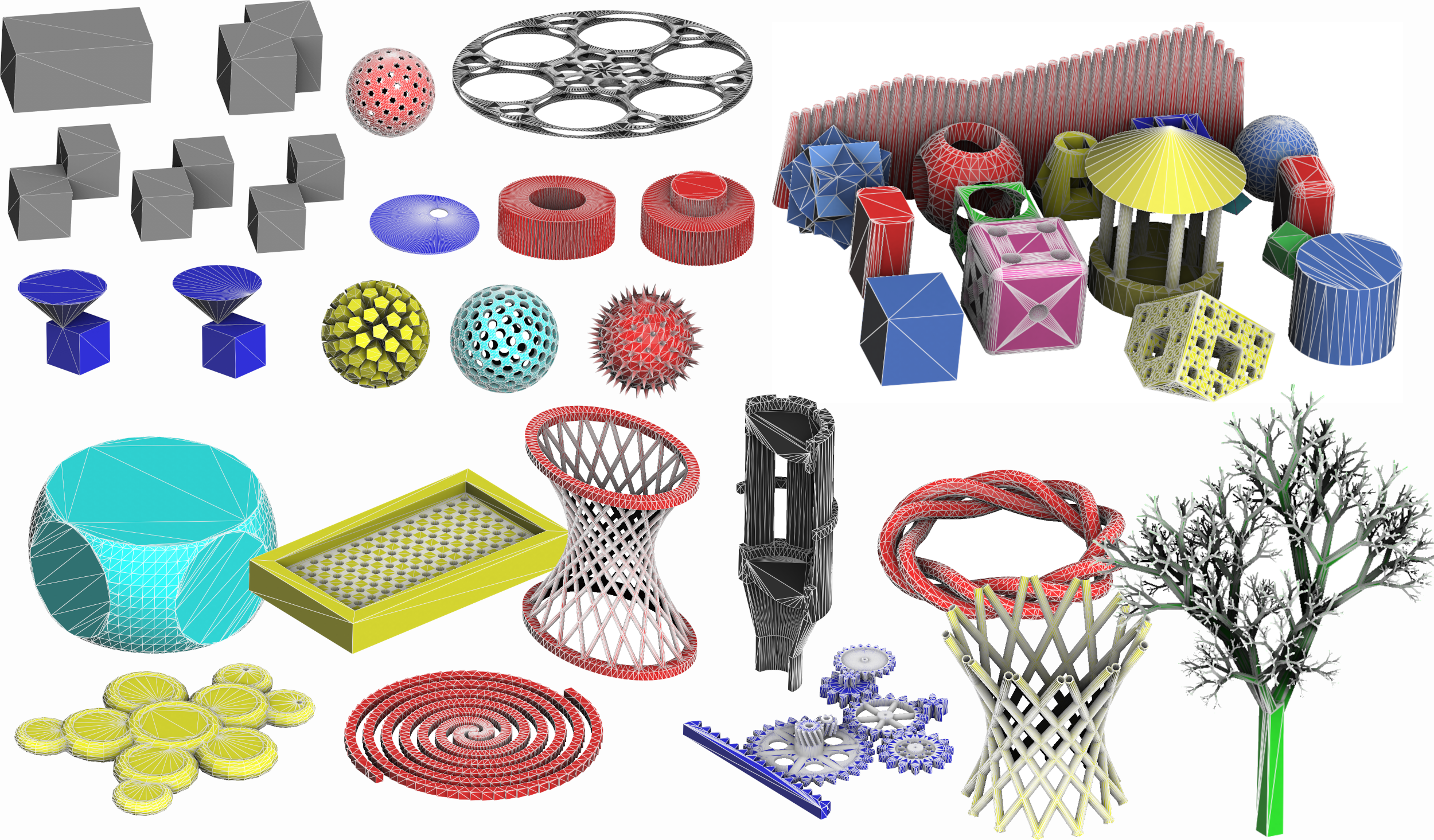}}
  \caption{ThingiCSG: a collection of openSCAD files from different sources}
  \label{fig:thingicsg}
\end{figure*}

\begin{figure}
  \centerline{\includegraphics[width=\columnwidth]{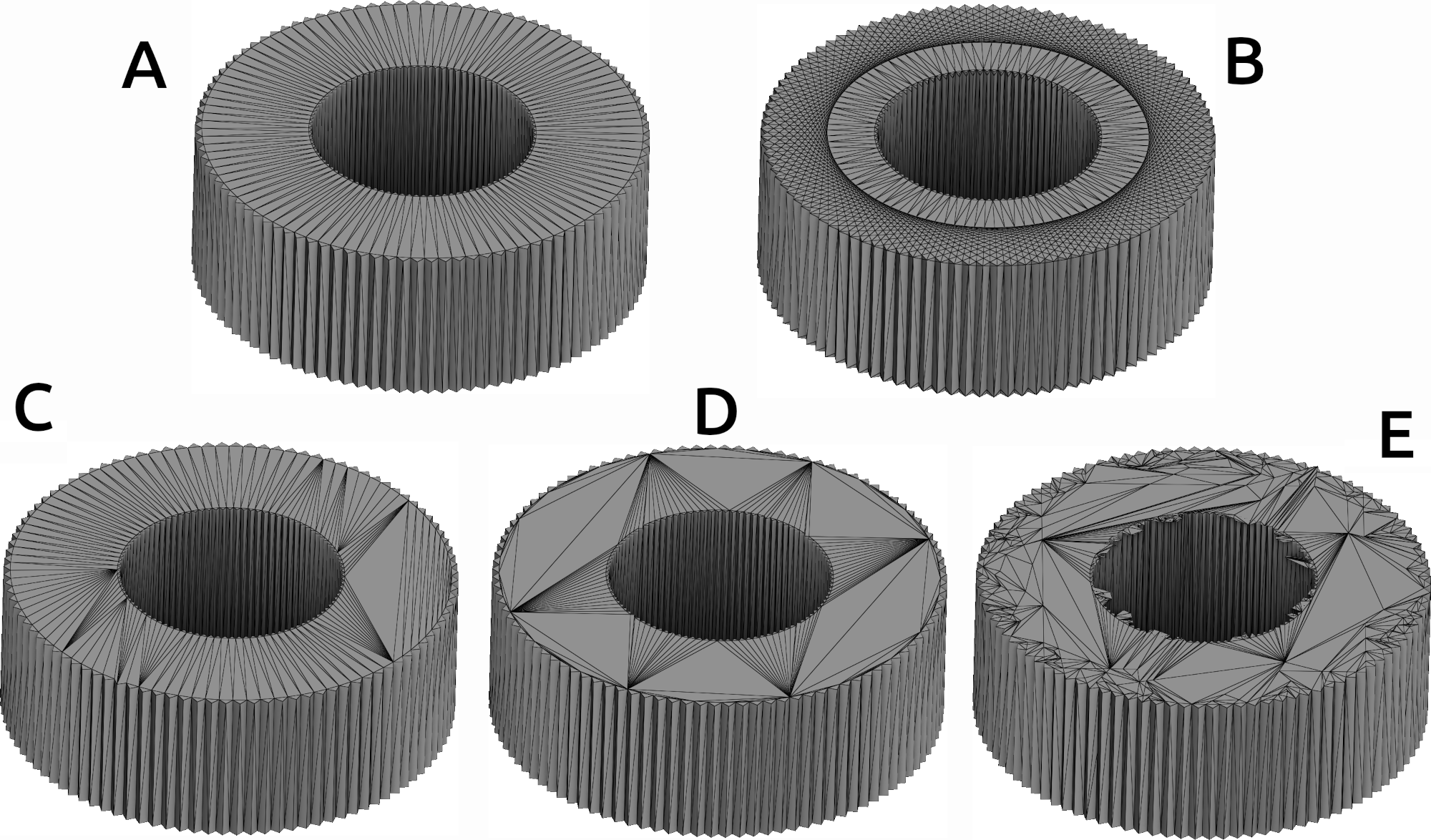}}
  \caption{ThingiCSG's {\tt nasty\_gears\_1} model, composed of the difference between two sets of
    50 rotated cubes. This creates many co-planar facets. A: our result: B: our result without
    simplification of co-planar facets; C: CGAL NEF result; D: CGAL corefinement result;
    E: ``manifold'' kernel result.}
  \label{fig:nasty_gear_1}
\end{figure}

\begin{table}
  \begin{tabular}{lllll}
    \hline
    ID & ours & CGAL & CGAL   & manifold \\
       &      & NEF  & coref. &          \\
    \hline
\verb|christmas_ball|    & 6  & 71  & 4   & 1 \\
\verb|cube_cone_1..2|    & <1  & <1   & <1   & <1 \\
\verb|demo_reel|         & 8  & 29  & 121 & 1 \\
\verb|demo_reel_u..|     & 5  & 18  & 117 & <1 \\
\verb|demo_reel_u.._n..| & 2  & 8   & 2   & <1 \\
\verb|fibo_cylinders|    & <1  & 4   & <1   & <1 \\
\verb|fishy_sphere|      & 45 & 513 & 17  & 1 \\
\verb|golf|              & 5  & 176 & 3   & 1 \\
\verb|hollow_ball_bunny| & 8  & 171 & 6   & 1 \\
\verb|hollow_ball|       & 3  & 50  & 3   & 1 \\
\verb|multi_rot_cube|    & 43 & 15  & 13  & <1 \\
\verb|nasty_gear_0|      & 2  & 9   & 2   & <1 \\
\verb|nasty_gear_1|      & 4  & 11  & 2   & <1 \\
\verb|nasty_gear_2|      & 5  & 9   & 2   & <1 \\
\verb|nasty_gear_3|      & 4  & 9   & 3   & <1 \\
\verb|nasty_gear_4|      & 4  & 10  & 2   & <1 \\
\verb|seven_sins_2|      & 2  & 9   & 24  & <1 \\
\verb|seven_sins_3|      & 3  & 19  & 235 & <1 \\
\verb|seven_sins_4|      & 3  & 20  & 244 & <1 \\
\verb|seven_sins|        & 2  & 6   & 1   & <1 \\
\verb|spiky|             & <1  & 14  & <1   & <1 \\
\verb|three_cubes|       & <1  & <1   & <1   & <1 \\
\verb|two_cubes_1..5|    & <1  & <1   & <1   & <1 \\
\verb|two_cylinders_1|   & 4  & 2   & 29  & <1 \\
\verb|two_cylinders_2|   & 1  & 1   & 1   & <1 \\
    \hline
  \end{tabular}
  \caption{Timings (in seconds) for the {\tt Basic} collection of ThingiCSG.}
  \label{table:thingicsg3}
\end{table}

\begin{table}
  \begin{tabular}{lllll}
    \hline
    ID & ours & CGAL   & manifold \\
       &      & coref. &          \\
    \hline
\verb|fibo_bunny_union| & 749  & 1482 & 380 \\
\verb|fibo_bunny_diff.| & 985  & 1450 & 311 \\
\verb|fibo_sphere_20|   & 3    & 5    & 1   \\
\verb|fibo_sphere_100|  & 44   & 69   & 18  \\
\verb|fibo_sphere_200|  & 200  & 292  & 72  \\
\verb|fibo_sphere_500|  & 2225 & X    & 385 \\
    \hline
  \end{tabular}
  \caption{Timings (in seconds) for the {\tt Large} collection of ThingiCSG.}
  \label{table:thingicsg4}
\end{table}

\subsection{ThingiCSG}
\label{sec:thingiCSG}

To test the Weiler model and classification algorithm, I collected 83 files
in the OpenSCAD format from different locations (some of them are displayed
in Figure \ref{fig:thingicsg}):
\begin{itemize}
  \item the OpenSCAD examples and test suite \cite{WEB:OpenSCAD}, with examples of increasing complexity;
  \item an OpenSCAD tutorial \cite{WEB:OpenSCADtuto}, with more complicated examples;
  \item the files from my non-regression test suite, with small but challenging examples, with degeneracies, as well as larger ones, such has the ``Fibobunny shere'' on the first page.
\end{itemize}

OpenSCAD has two different file formats: the \verb|.scad| format, that corresponds
to a complete programming language, and the \verb|.csg| format, limited to a
subset of the OpenSCAD language, corresponding to ``flat CSG trees'',
with only primitive and CSG operations.
I implemented a parser for the \verb|.csg| format, that makes it easier to test
CSG operations (an alternative would have been to implement a backend for OpenSCAD).
One can use OpenSCAD to convert from the \verb|.scad| to the \verb|.csg| file format.

Both OpenSCAD files and the \verb|.csg| parser are available in a new
\verb|thingiCSG| repository, to make it easy to test and benchmark new research
projects on mesh CSG.

I shall now give some statistics and comparisons, using:
\begin{itemize}
\item the algorithm presented in this article.
\item the default OpenSCAD backend, based on CGAL NEF polyhedra \cite{NefComplexes},
\item the OpenSCAD backend, based on CGAL co-refinement \cite{WEB:CGALcorefinement},
\item the ``manifold'' OpenSCAD backend \cite{WEB:Manifold}, based on \cite{ManifoldArticle}
\end{itemize}

On the OpenSCAD examples collection (Figure \ref{fig:thingicsg} top right),
that has 29 files, the four kernels take 1s and less
on each file, except the NEF kernel, that takes 28s on \verb|example006.scad| and \verb|example024.scad|,
and that takes a few seconds on \verb|example010.scad|, \verb|example018.scad| and above.

The statistics for the four kernel on the \verb|Presentation| collection (Figure \ref{fig:thingicsg} bottom) are reported in Table \ref{table:thingicsg2}, and the statistics for the \verb|Basic| collection, with custom small-but-challenging models that I created,
are shown in Table \ref{table:thingicsg3}. As can be seen, the ``manifold'' kernel is
always the fastest. Our method is often faster than the co-refinement kernel, except in a small
number of cases. A visual comparison on one of the examples (\verb|nasty_gears_1|) is shown
in Figure \ref{fig:nasty_gear_1}. This example is challenging, because it is made of the
difference between two sets of 50 cubes rotated around their axis. It generates a very large
number of co-planar facets, stressing both the arithmetic kernel and the constrained Delaunay
triangulation. Our result is shown in Figure \ref{fig:nasty_gear_1}-A (with co-planar facets
simplification) and B (without co-planar simplification). The two CGAL-based kernels (C and D)
produce a correct result (but it is not a Delaunay triangulation), and the ``manifold'' kernel
fails producing a correct result for this specific example, as well as other ones with
similar co-planar configurations or high mesh density.

In table \ref{table:thingicsg4}, timings are reported for larger datasets, such as the union and difference of sphere with
200 Stanford bunnies with a Fibonacci distribution displayed on the first page. Each bunny has 75K vertices. The table also
reports timings for the union of 200 spheres of various resolution (between 200 and 125K vertices). For the largest example,
\verb|fibo_sphere_500|, where the input has 62.5M vertices, the CGAL corefinement kernel crashed (out of memory). The ``manifold''
kernel did output a result in 385s, but this result has many missing triangles.

\section{Conclusions}
\label{sec:conclusions}

To conclude this article, I shall report some lessons I learned from this project, as well as
possible directions for future work. What I find important to remember about this experience is the
following list. Most elements will probably be not a big surprise, but I found it worth mentioning them:

\begin{itemize}
  \item Non-regression testing is, as always, extremely important, especially for this algorithm, that
    has many parts, and each part is complicated;
  \item assertion checks everywhere in the code (e.g., testing the combinatorial consistency of the 3-Map,
    the Delaunay property of the triangulation) helps a lot detecting bugs at an early stage;
  \item Thingi10K is interesting, with such a large database, many corner cases are likely to happen,
    such as projection axis requiring exact precision \verb|#356074|, 3D grid of mutually intersecting
    triangles that generate a huge number of intersections \verb|#996816| stressing the constrained
    Delaunay triangulation code, or mesh with a large number of very skinny triangles mutually intersecting
    around a pole \verb|#101633| stressing the arithmetic kernel. Solving these issues helped identifying
    hot spots that would have remained hidden otherwise;
  \item the arithmetic kernel can be stressed a lot by some meshes. Carefully optimizing the arithmetic kernel
    can be a key for optimal performance. The ``complexity of the coordinates'' impacts performance a lot,
    in particular with arithmetic expansions, but not only. For the predicates, use expressions that
    are as simple as possible, and compress often;
  \item a predicate can have different equivalent expressions. One can use one of them for the filter, another one
    for the exact evaluation and a third one for the symbolic perturbation;
  \item arithmetic expansions can be pushed a little bit, but soon one reaches their two limitations: (1) with cascaded
    constructions / predicates, exponents can overflow; (2) operations start to cost a lot with expansions longer than
    a few tenths of components. As soon as constructions are involved, multi-precision seems to be a better solution;
  \item implementing some well known algorithms, such as a 2D constrained Delaunay triangulation with intersecting constraints is delicate.
    Often, the main algorithm is simple, just as in the textbook, but most of the textbooks and existing references make simplifying
    assumptions (e.g., points in generic positions). For instance, in a 2D constrained Delaunay triangulation, detecting the edges
    that intersect a constraint with all the possible corner cases much subtler than one would think in the first place;
  \item exact constructions combined with exact predicates with interval filters appears to be a reasonable option for mesh intersection,
    which I was not sure when starting this project. This is probably because in a (not too convoluted) mesh with $N$ elements,
    there are approximately $\sqrt{N}$ elements in the intersection. One can afford paying more for these elements, because in general
    their number does not grow too quickly in function of $N$ (except for \verb|#996816| of course!);
  \item the ``simply implement the math'' vision is nearly achievable. One can write elegant code that looks like the textbook
    algorithm, but this comes at the expense of carefully optimizing and treating all the corner cases in the arithmetic kernel,
    constructions and predicates. From a software
    engineering point of view, doing so is interesting because the most complicated code is
    confined into easy-to-test functions with a well defined API.
\end{itemize}

This work can be improved and extended in several directions. Clearly, the main missing component is a ``snap rounding'' algorithm, that
transforms the exact coordinates into floating point numbers while preserving some topological properties. The approach described
in \cite{devillers_et_al:LIPIcs:2018:8743,valque:hal-02393625} is very promising. Another interesting direction is the approach
completely based on floating-point numbers described in \cite{ManifoldArticle,WEB:Manifold}. It is a completely different paradigm.
While the ``manifold'' OpenSCAD kernel, based on this paradigm, does not always output a correct result, it does very often,
at a spectacular speed. It is very rare that it takes more than 1s for a CSG. Is it possible to derive an algorithm with the
same performance and more guarantees? \new{It is also important to mention here that if using integer-only coordinates
  is allowed by the application context, a completely different class of methods
  can be used, such as the ones presented in \cite{10.1145/3528223.3530181,NehringWirxel2021FastEB}. They result
  in a spectacular acceleration factor (50x to 100x) as compared to what is presented here. }
Finally, it may be interesting to enrich the ThingiCSG database introduced in this article with a larger set of examples, especially if a larger research community wants to tackle this type of problems.

\begin{acks}
  I wish to thank Nicolas Ray for many discussions and for vigorously encouraging me to explore the
  simplest option first. This research was supported by the Inria AeX COSMOGRAM-Launchpad grant.
\end{acks}

\bibliographystyle{ACM-Reference-Format}
\bibliography{CSG}

\appendix
\section{Online Resources}

\begin{itemize}
\item The main algorithm, expansion-based kernel and OpenSCAD CSG parser are available in the GEOGRAM library:\\ \url{https://github.com/BrunoLevy/geogram}
\item The ThingiCSG collection of CSG trees in OpenSCAD format is available here: \url{https://github.com/BrunoLevy/thingiCSG}
\item Thingi10K \cite{Thingi10K} is available here: \\
  \url{https://ten-thousand-models.appspot.com/},
\end{itemize}

\end{document}